\documentclass[prd,showkeys]{revtex4}
	\usepackage{amsmath,amssymb}
	\usepackage{enumerate}
	\usepackage{amsfonts}
	\usepackage[usenames,dvipsnames]{pstricks}
	\usepackage{subfig}
	\usepackage{epsfig}
	\usepackage[colorlinks,hyperindex]{hyperref}
	\hypersetup
	{
		colorlinks,%
		citecolor=black,%
		linkcolor=black,%
		urlcolor=black,%
	}
\usepackage{allpurpose}
\usepackage{doi}
\usepackage{tikz}
\usetikzlibrary{calc,patterns,angles,quotes}
\usepackage{multirow}
\usepackage{makecell} 
\usepackage{hhline} 

\begin{document}

\title{Gravitational lensing of massive particles in Reissner-Nordstr\"om spacetime}
\date{\today}

\author{Xiankai Pang}
\affiliation{Center for Theoretical Physics, School of Physics and Technology, Wuhan University, Wuhan, 430072, China}

\author{Junji Jia}
\email[Email: ]{junjijia@whu.edu.cn}
\affiliation{MOE Key Laboratory of Artificial Micro- and Nano-structures, School of Physics and Technology,  Wuhan University, 430072, China}
\affiliation{Center for Astrophysics, School of Physics and Technology,  Wuhan University, Wuhan, 430072, China}

\begin{abstract}
  In this work we study the deflection angle $\Delta \varphi$ and gravitational lensing of both lightlike and timelike neutral rays in \RN (RN) spacetimes. The exact deflection angle is found as an elliptical function of the impact parameter $b$ and velocity $v$ of the ray, and the charge $Q$ of the spacetime. In obtaining this angle, we found the critical impact parameter $b_c$ and radius of particle sphere $r_c$ that are also dependent on $v$ and $Q$. In general, both the increase of velocity and charge reduces the $b_c$ as well as $r_c$. To study the effect of $v$ and $Q$ on the deflection angle $\Delta\varphi$, its weak and strong deflection limits, relativistic and non-relativistic limits, and small charge and extremal RN limits are analyzed carefully. It is found that both the increase of velocity and charge reduces the deflection angle. For weak deflection, the velocity and charge corrections appear respectively in the $\mathcal{O}(1/b)$ and $\mathcal{O}(1/b^2)$ orders. For strong deflections, these two corrections appear in the same order. The apparent angles and magnifications of weak and strong regular lensing, and retro-lensing are studied for both lightlike and timelike rays. In general, in all cases the increase of velocity or charge will decrease the apparent angle of any order.  We show that velocity correction is much larger than that of charge in the weak lensing case, while their effects in the strong regular lensing and retro-lensing are comparable. It is further shown that the apparent angle and magnification in strong regular lensing and retro-lensing can be effectively unified. These observables at different orders in these two kinds of lensing are staggered: the apparent angles can be ordered in a staggered way and the magnifications forms two staggered geometric series. Finally, we argue that the correction of $v$ and $Q$ on the apparent angle can be correlated to mass or mass hierarchy of timelike particles with certain energy. In addition, the effects of $v$ and $Q$ on shadow size of black holes are discussed.
\end{abstract}

\keywords{Deflection angle; Gravitational lensing; Charged black hole; Strong field limit}

\maketitle

\section{Introduction}

The deflection of light ray was one of the most famous prediction of classical general relativity  (GR)  \cite{Einstein:1911vc}. Its confirmation in 1919  \cite{Dyson:1920cwa} not only helped establishing GR as a correct theory describing gravity, but also laid the foundation of its usage in astrophysics and cosmology. The gravitational lensing  (GL) effect based on the light ray deflection, has become one of the most important tools in measuring properties such as mass of galaxies or clusters, determining Hubble constant, and studying properties of distant galaxies. Lately, strong  (or macro) GL has been used to constrain dark energy and the density profile of the lensing galaxy   \cite{Walsh:1979nx,Oguri:2010ns,Treu:2010uj}. In more recent years, microlensing has been used to study the dark objects in the galaxy halos   \cite{Aubourg:1993wb,Alcock:2000ph}
and detection of extrasolar planets   \cite{Gaudi:2008zq,Gould:2010bk}. In the past few years, the GL of supernova by clusters and single galaxy have also been detected   \cite{Nordin:2013cfa,Quimby:2013lfa,Kelly:2014mwa,Goobar:2016uuf}.

On the other hand, it is well known that in a supernova the vast majority of the energy is released through the emissions of neutrinos. These neutrinos have been detected in SN1987A and are known to have non-zero masses. Therefore, in principle these neutrino flux will also reach the observatory and might be gravitationally lensed. The detection of gravitational wave  (GW) together with its electromagnetic counterparts   \cite{TheLIGOScientific:2017qsa, GBM:2017lvd, Monitor:2017mdv} also inspired works using the GL of GWs to measure Hubble constant   \cite{Liao:2017ioi}.   Detection of lensing of such particles will not only reveal properties of the source  (e.g. supernova mechanism) and the lens  (e.g. mass, charge and angular momentum), but also the properties of these particles themselves. In the case of the neutrino, its absolute mass and mass hierarchy can be related to the deflection angle of the trajectory   \cite{Jia:2015zon}. In the case of GW, its speed has already been severely constrained by the small time-delay between GW and its electromagnetic counterpart and potentially more so by its GL. All these applications however require a better understanding of the influence of particle velocity on the deflection of the trajectory and GL, in addition to the effect of other properties of the lens, such as its charge and angular momentum.

In this work, we plan to study the deflection angle and GL of both lightlike and timelike neutral rays in RN spacetimes.
Previously, trajectory in RN spacetime have been classified in Ref.   \cite{Chandrasekhar:1985kt, Grunau:2010gd, Hackmann:2008tu, Pugliese:2011py}. Sereno   \cite{Sereno:2003nd} and Keeton and Petters   \cite{Keeton:2005jd} studied in the weak field limit, and Eiroa in the strong field limit and numerically   \cite{Eiroa:2002mk}, the deflection angle and GL of light ray in RN spacetimes. Bozza   \cite[Eq.~(35)]{Bozza:2002zj} proposed a general formula for the deflection of light ray in strong field limit in static and spherically symmetric spacetimes, while Amore et al.   \cite{Amore:2006mc} developed an nonperturbative way of approximating the deflection angles in both strong and weak field limits. Both these two methods are applied to the case of RN metric. Bin-Nun studied how the charge in galactic center  (supposed a RN metric) would affect the relativistic images  \cite{BinNun:2010ty}, which are formed after the first image on each side of the optical axis \cite{Virbhadra:1999nm}. Zakharov proposed to use the shadow size to constraint charge in the galactic center   \cite{Zakharov:2011zz, Zakharov:2014lqa}. More recently, Tsukamoto and Gong studied the retro-lensing of RN black hole for general charge   \cite{Tsukamoto:2016oca}. However, all these studies concentrated on {\it either} of the weak or strong field limit for the deflection and GL, and most importantly {\it only} for lightlike rays. A thorough study of the deflection angle and observables in various lensing scenarios, including weak, strong and retro-lensing, of particles with general velocity in RN spacetime, is still missing.

In this work, we will derive a general formula for the deflection angle in RN spacetime for arbitrary velocity. This angle is then studied in various limits, including the weak and strong field limits, relativistic and non-relativistic particle limits and small charge and extremal black hole limits. Under these limits, we work out the apparent angles and magnifications for the weak, strong and retro-lensing cases. We pay special attention to the influence of velocity of the ray and the charge of the spacetimes on these observables.

The work is organized as the following. In section \ref{sec:deflectionangle} we first derive the equation of particle trajectories in an integral form, and then give the critical impact parameter and particle sphere radius and analyze their various special cases. In the end of this section we do the integral and obtain an exact deflection angle in the form of the first kind of incomplete elliptic function. In section \ref{sec:weakstrong} we expand this exact formula for both relativistic and non-relativistic particles in weak and strong field limits. It will be shown that both the particle velocity increase and charge increase, will reduce the deflection angle, although orders of effect of these two parameters are different. In section \ref{sec:apparentangle} we obtain the apparent angles and magnifications of weakly, strongly and retro- lensed light and relativistic particles rays. The effect of ray velocity and charge are analyzed. Features of the apparent angles and magnifications of strong regularly lensed and retro-lensed images, and their inter-relations are discussed. Finally, in section \ref{sec:discussions} we discuss some potential extensions and applications of these results.

Through the paper we use the units such that $G=c=4\pi\varepsilon_0=1$, where $G$ is the gravity constant, $c$ is the speed of light and $\varepsilon_0$ is the vacuum permittivity.

\section{Deflection angle, particle sphere and critical impact parameter} \label{sec:deflectionangle}

We start with the RN metric in its conventional form
\iea{
  \dd s^2=f(r)\dd t^2-\frac{1}{f(r)}\dd r^2-r^2\dd \Omega^2,\label{eq:rnmetric}
}
where $\displaystyle \dd \Omega^2=\dd \theta^2+\sin^2 \theta \dd \varphi^2$ is the solid angle and
\iea{
  \displaystyle f(r)=1-\frac{2M}{r}+\frac{Q^2}{r^2},
}
with \mcm and \mcq being mass and electrical charge of the central body. The RN black hole has two horizons with radius
\iea{
  r_\pm=M\pm\sqrt{M^2-Q^2}. \label{eq:rhorizon}
}
In this work we assume $0\leq Q\leq M$, so that at least one horizon exists. Because we will study the trajectory of neutral particles, the sign of $Q$ doesn't matter.

Due to the spherical and static symmetry of the RN metric \eqref{eq:rnmetric}, the particle will move in a plane which can be chosen as the one with $\theta=\pi/2$. Furthermore, the energy $E$ and angular momentum $L$ per unit mass are also conserved in this spacetime, and therefore they can be expressed by velocity and impact parameter $b$ at infinity where the spacetime is flat as
\iea{
  E=\frac{1}{\sqrt{1-v^2}},\quad L=\frac{v}{\sqrt{1-v^2}}b. \label{eq:eldef}
}
For massless particles, $L/E=bv=b$ holds.
The particle motion can be found by using the following action \cite{Landau:1982dva}
\iea{
  S=-Et+L\varphi+S_r(r,~L,~E), \label{eq:action}
}
where by the virture of energy and angular momentum conservation, we have already sperated the variables $t$ and $\varphi$. Here $S_r(r,~L,~E)$ is an unknown function independant of $t$ and $\varphi$ and to be determined by the Hamilton-Jacobi equation: 
\iea{
  \frac{E^2}{f(r)}-\frac{L^2}{r^2}-f(r)\left(\frac{\dd S_r(r)}{\dd r}\right)^2=\kappa,\label{eq:hj}
}
where $\kappa=0,~1$ for massless and massive particle respectively.
Solving $\displaystyle S_r(r,E,L)$ from Eq. \eqref{eq:hj} and substituting back into Eq. \eqref{eq:action}, we can obtain the action. From that, the orbital equation of $\varphi$ in terms of the radius $r$ is found to be \cite{Landau:1982dva}
\iea{
  \frac{\p S}{\p L}=\mathrm{constant}. \label{eq:pspL}
}
Further solving this,  we obtain
\iea{
  \varphi(r)=\int\frac{L}{r^2}\frac{\dd r}{\sqrt{E^2-V(r)}},  \label{eq:phir}
}
with the effective potential
\iea{
  V(r)=\left(1-\frac{2M}{r}+\frac{Q^2}{r^2}\right)\left(\frac{L^2}{r^2}+\kappa\right).\label{eq:effectiveV}
}
The physical motion only occurs in the domain $E^2\geq V^2(r)$. Note that Eq. \eqref{eq:phir} is quite general and can be used in the study of bounded trajectories as well. Defining the following dimensionless variables
\iea{
  \omega=\frac{M}{r},~f=\frac{bv}{M},~h=\frac{Q}{M},\label{eq:rel1}
}
Eq. \eqref{eq:phir} can be cast into the following form
\iea{
  \varphi(\omega)=\int\frac{f\dd \omega}{\sqrt{\left[\omega^2f^2+(1-v^2)\kappa\right]\left[(2-h^2\omega)\omega-1\right]+1}}. \label{eq:phiomega}
}
Even though the physical meaning of $f$ is not as apparent as the impact parameter $b$, we will see that $f$ is easier to handle than the impact parameter $b$ for non-relativistic particles and equals $b$ for relativistic particles.

For massive particles we set $\kappa=1$ in Eq. \eqref{eq:phiomega}
\iea{
  \varphi(\omega)= \int\frac{f\dd \omega}{\sqrt{\left[\omega^2f^2+(1-v^2)\right]\left[(2-h^2\omega)\omega-1\right]+1}}. \label{eq:phiomegamass}
}
By letting $v=1$ in Eq. \eqref{eq:phiomegamass} or $\kappa=0$ in Eq. \eqref{eq:phiomega} we get the same result for massless particles
\iea{
  \varphi(\omega)= \int\frac{f\dd \omega}{\sqrt{\omega^2f^2\left[(2-h^2\omega)\omega-1\right]+1}}. \label{eq:phiomegalight}
}
This means that $\varphi(\omega)$ is continuous when the velocity of the particle changes from relativistic to exactly the speed of light. This also implies that the deflection angle of relativistic massive particles will always be close to that of light.

Particles traveling from a source at coordinate $(r_i,~\varphi_i)$ to a detector at coordinate ($r_f,~\varphi_f$), experiences a deflection angle $\Delta\varphi=\varphi_f-\varphi_i$.  Usually the source and the detector are far away from the center and thus $\omega_i=M/r_i$ and $\omega_f=M/r_f$ are set to zero. Specifying the orbital Eq. \eqref{eq:phiomegamass} to unbounded orbits yields the deflection angle
\iea{
  \Delta\varphi=2\int_0^{\omega_{2}}\frac{f\dd \omega}{\sqrt{-f^2h^2(\omega-\omega_1)(\omega-\omega_2)(\omega-\omega_3)(\omega-\omega_4)}}, \label{eq:deltaphiint}
}
where $\omega_1<0<\omega_2<\omega_3<\omega_4$ are four real roots of the quartic equation in the denominator of \eqref{eq:deltaphiint}
\iea{
  \left[\omega^2f^2+(1-v^2)\right]\left[(2-h^2\omega)\omega-1\right]+1=0. \label{eq:polyomega}
}

To ensure the particle will return back to spatial infinity without entering into horizons, all four roots have to be real. This demands that the discriminant $\Delta$ of Eq. \eqref{eq:polyomega} (see Eq. \eqref{eq:quarticdiscri}) is positive. From this, we obtain a critical value $f_c$ in terms of $h$ and $v$
\iea{
  f_c(h,v)&=& \frac{\sqrt{1-v^2} \sqrt{h^2 \omega _c(h,v)-1}}{\sqrt{-2 h^2 \omega _c(h,v)^3+3 \omega _c(h,v)^2-\omega _c(h,v)}}, \label{eq:fcsolomegac}
}
where $\omega_c(h,v)$ is also a critical value given by $\omega_2$ in Eq. \eqref{eq:quarticsol23} with coefficients in Eq. \eqref{eq:omegacabcde}. Although the expression of $\omega_c$ is long and we only list it in the appendix \ref{sec:exact}, the fact that it is an elementary function of parameters $h$ and $v$ is clear. The $f_c(h,v)$ in turn determines a critical impact parameter $b_c(h,v)$ through Eq. \eqref{eq:rel1}
\be b_c(h,v)=\frac{M f_c(h,v)}{v}. \label{eq:defbc}\ee
And the critical $\omega_c(h,v)$ defines a critical closest radius
\iea{r_c(h,v)=\frac{M}{\omega_c(h,v)}.\label{eq:defrc}}
Particles with impact parameters smaller than $b_c$ will enter the region with radius $r_c(h,v)$, and then be captured. Particle with impact parameters larger than $b_c$ will have a closest distance larger than $r_c$ and then escape back to infinity under geodesic motion. Therefore, $r_c$ defines an sphere that is analogous to the photon sphere in the case of photon geodesic motion in the spacetime and can be called {\it particle sphere} of the RN spacetime. For an precise definition of photon sphere and its generalization to photon surface, see Ref. \cite{Claudel:2000yi}.
\begin{figure}[htp]
  \centering
  \includegraphics[width=0.45\textwidth]{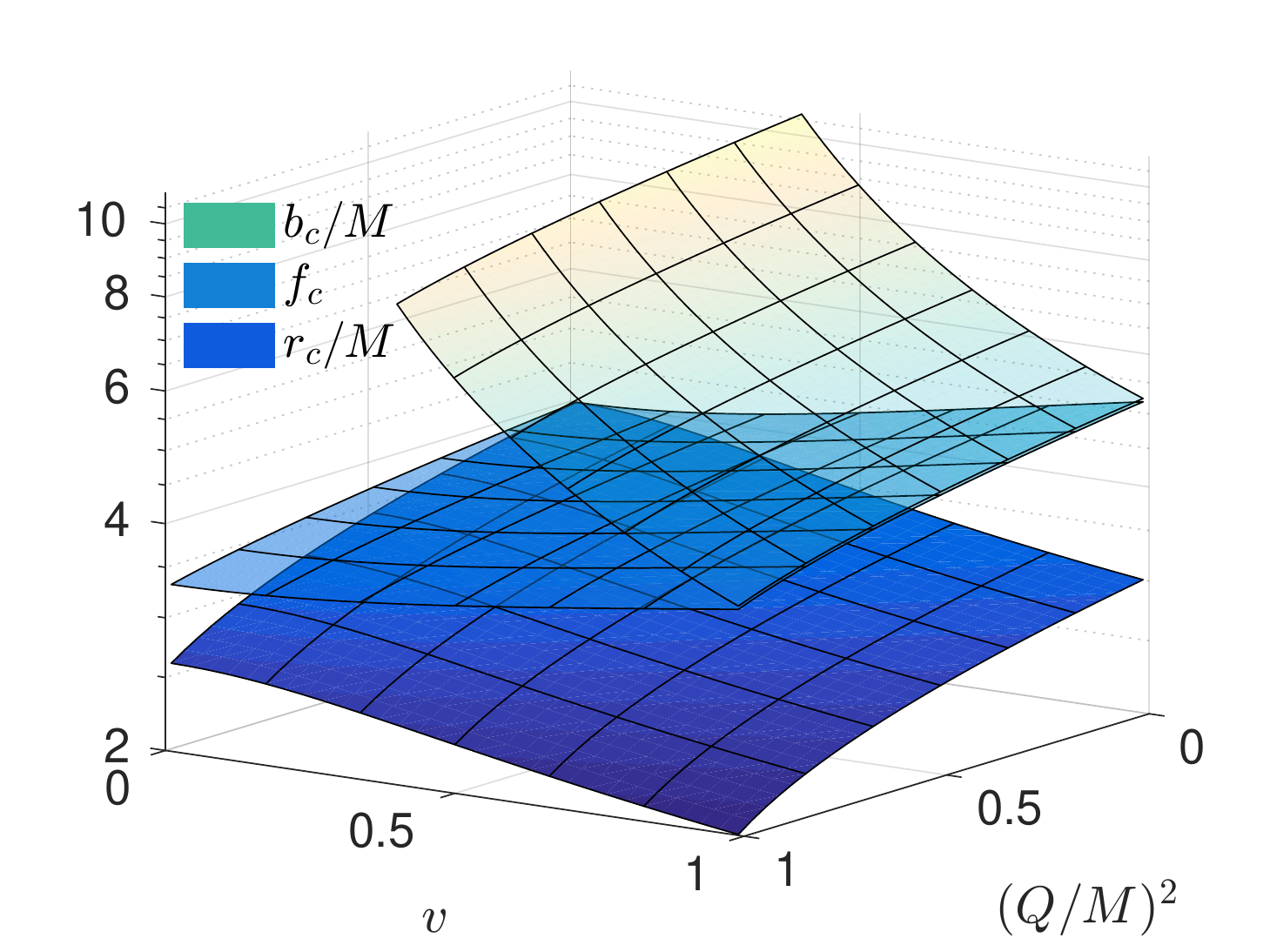}
  \caption{The critical impact parameter $b_c$, the critical ratio $f_c$ and the radius of particle sphere $r_c=M/\omega_c$.}
  \label{fig:rcfcbc}
\end{figure}

In Fig. \ref{fig:rcfcbc}, we plot the  particle sphere radius $r_c/M$, critical $f_c$ and critical impact parameter $b_c/M$ as functions of $h$ and $v$.
It can be seen that $r_c$ range from $2M$ to $4M$ in the entire parameter space of $h$ and $v$. In the Schwarzschild case, $r_c$  takes the form
\iea{
  r_c(h=0,v)&=&M\left(2+\frac{4}{\sqrt{8v^2+1}+1}\right).
}
Therefore, it increases from $3M$ to $4M$ as velocity decreases from $1$ to $0$.
And the critical impact parameter $b_c$ in this case is given by
\iea{
  b_c(h=0,v)&=& \frac{M\left[8v^4+20v^2-1+\left(8v^2+1\right)^{3/2}\right]^{1/2}}{\sqrt{2}v^2}.
}
For extremal RN spacetime, $r_c$ and $b_c$ become respectively
\iea{
  r_c(h=1,v)&=&\frac{M \sqrt{3(1- v^2)}}{\sqrt{3(1- v^2)}-2 \sqrt{2} \sin \left( \alpha\right)}, \label{eq:rch0} \\
  b_c(h=1,v)&=& \frac{\sqrt{3} M \left(1-v^2\right)}{v \sqrt{-11+3 v^2+6 \sqrt{6(1- v^2)} \sin \left( \alpha\right)+8 \cos \left(2 \alpha\right)}},\label{eq:bch0} 
}
where
\iea{
  \alpha&=& \frac{1}{3}\sin^{-1}\left[\frac{3}{4\sqrt{2}} \sqrt{3(1-v^2)}\right].
}
It is clear that $r_c$ increase from $2M$ to $2M/(3-\sqrt{5})$ as $v$ decrease from $1$ to $0$. For $h=1,~v=1$ we see form Fig. \ref{fig:rcfcbc} that $r_c=2M$, and circular orbits in this sphere are stable  \cite{Khoo:2016xqv}.

For any given $v$, increase of  charge $Q$ reduces the radius of particle sphere. For lightlike rays in particular, the photon sphere radius and the critical impact parameter are
\iea{
  r_c(h,v=1)&=&\frac{4 h^2 M}{3-\sqrt{9-8 h^2}},\label{rchv1}\\
  b_c(h,v=1)&=& M\sqrt{\frac{4 h^2(2h^2-9)+\left(9-8 h^2\right)^{3/2}+27}{2(1- h^2)}}.\label{eq:bchv1}
}
Note that for Eq. \eqref{rchv1}, the $r_c=3M$ limit of the Schwarzschild spacetime will be recovered as $h$ approaches zero. From Eq. \eqref{eq:bchv1} we see that the $b_c$ for lightlike rays decreases from $3\sqrt{3}M$ for $Q=0$ to $4M$ for extremal RN spacetime.
On the other hand, the radius of particle sphere for $v=0$ is given by
\iea{
  r_c(h,v=0)&=&\frac{6M h^2}{6-2 \sqrt{3} \sqrt{\gamma }-\sqrt{36-12 \gamma -\frac{9 \sqrt{3} h^2}{\sqrt{\gamma }}}}, \label{eq:rcv0}
}
where
\iea{
  \gamma&=&1-\sqrt{4-3h^2}\cos \left\{\frac{1}{3} \cos ^{-1}\left[-\frac{27 h^4-144 h^2+128}{16 \left(4-3 h^2\right)^{3/2}}\right]\right\}.
}
It is seen from Eq. \eqref{rchv1} that for lightlike particles, $r_c$ decreases from $3M$ in the Schwarzschild $Q=0$ case to $2M$ in the extremal RN case, and from Eq. \eqref{eq:rcv0} then for $v$ approaching zero, $r_c$ decreases from $4M$ in the Schwarzschild case to $2M/ (3-\sqrt{5})$ in the extremal RN case.

 For the parameters that are within the critical values set above, we have solved the quartic equation \eqref{eq:polyomega} for the explicit form of its roots. However, these exact  roots are long functions of $h,~f$ and $v$ and are barely referred to in the main text except their approximations in various limits. Therefore, we only present their full form in appendix \ref{sec:exact} for reference. With the roots known, we can carry out the integration of Eq. \eqref{eq:deltaphiint}. The final result of the deflection angle, as a function of only $h,~f$ and $v$, then is \cite{Gradshteyn:1996table}
\iea{
  \Delta\varphi(h,f,v)&=& \frac{4 \ii F\left(\left.\sin ^{-1}\left(\sqrt{\frac{\omega _2 \left(\omega _1-\omega _3\right)}{\omega _3\left(\omega _1-\omega _2\right) }}\right)\right|\frac{\left(\omega _1-\omega _2\right) \left(\omega _3-\omega _4\right)}{\left(\omega _1-\omega _3\right) \left(\omega _2-\omega _4\right)}\right)}{h \sqrt{\left(\omega _1-\omega _3\right) \left(\omega _4-\omega _2\right)}}.\label{eq:deltaphi}
}
where  $F$ is the incomplete elliptic function following the convention of \emph{Mathematica} (see appendix \ref{sec:elliptic}).

In Fig. \ref{fig:deltaphi} we plot this deflection angle as a function of $b_M=b/M$ and $v$ for fixed $h=0$, and as a function of $Q$ and $v$ for fixed $b/M=7$ and $b/M=20$. It is clear from the plots that for any fixed $v$, in the weak field limit where $b$ is very large, the deflection angle approaches $\pi$. While in the strong field limit where $b$ approaches the critical value $b_c$, the deflection angle increases as $b$ decreases and eventually exceed $2\pi$ or any finite value. This corresponds to the trajectories that loop around the black hole once or many times. Moreover,
from Fig. \ref{fig:deltaphibmanydiffv} one finds that for fixed $b$, as the charge $h$ or velocity $v$ increase, the deflection angle decreases. This is due to the fact that as $h$ or $v$ increase, both $r_c$ and $b_c$ decrease. Therefore, trajectory with constant $b$ becomes effectively further away from the critical value and experiences less deflection. Furthermore, in the large $b$ limit Fig. \ref{fig:deltaphibmanydiffv} (see the $b/M=20$ plot) shows that the effect of charge to the deflection angle is much smaller than that of velocity. We will see in section \ref{sec:deflectionweak} that this is because in the large $b$ limit the charge correction appeared in the order of $\mathcal{O}(1/b^2)$ while the velocity effect appears at the order of $\mathcal{O}(1/b)$. While in the small $b$ limit (see the $b/M=7$ plot), the effects of both charge and velocity to the deflection angle are comparable. Again, we will show in section \ref{sec:deflectionweak} that in the critical $b$ limit, there effects are of the same order.

\begin{figure}[htp]
  \centering
  \subfloat[]{\label{fig:deltaphih0}\includegraphics[width=0.45\textwidth]{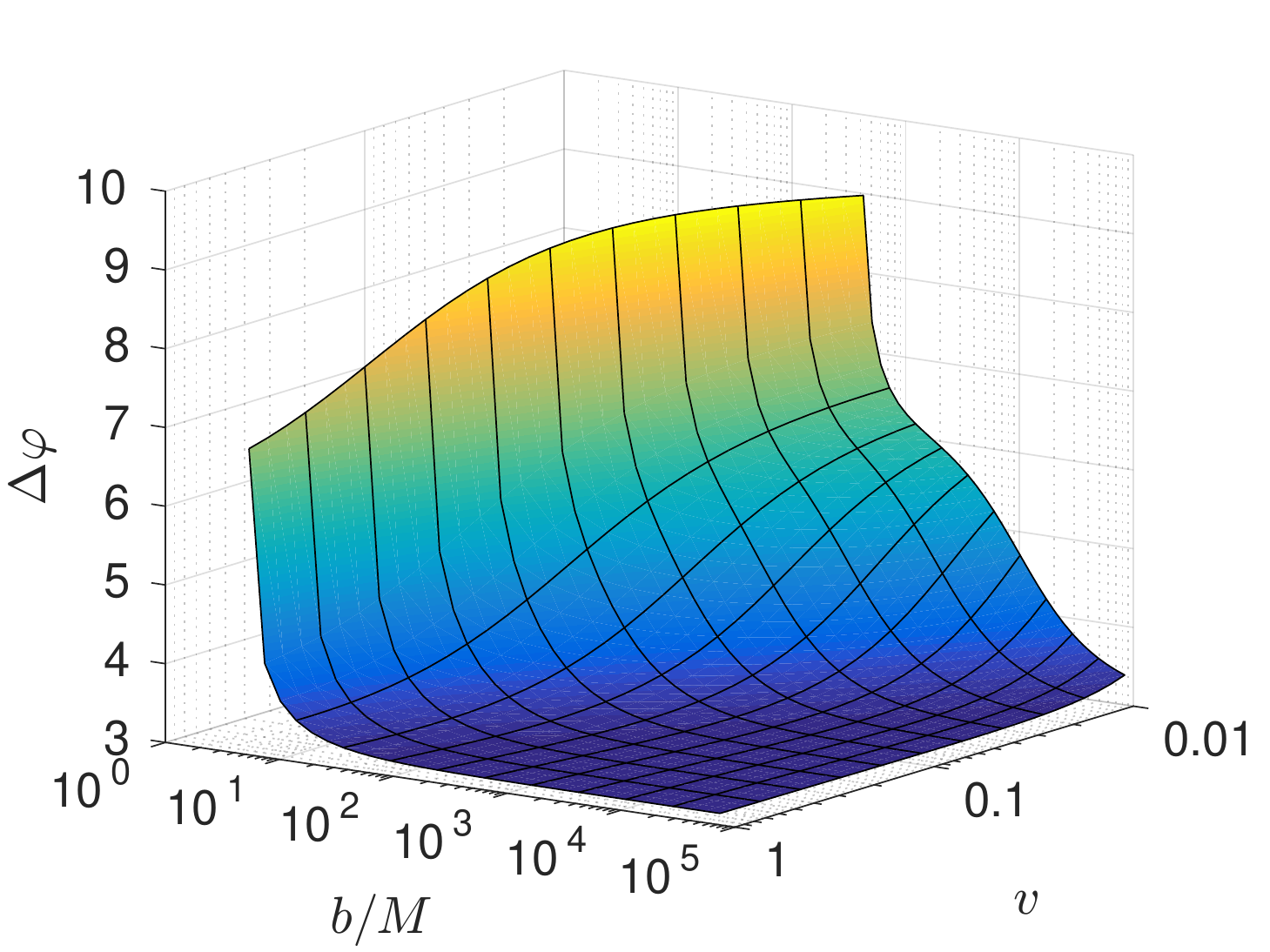}}~
  \subfloat[]{\label{fig:deltaphibmanydiffv}\includegraphics[width=0.45\textwidth]{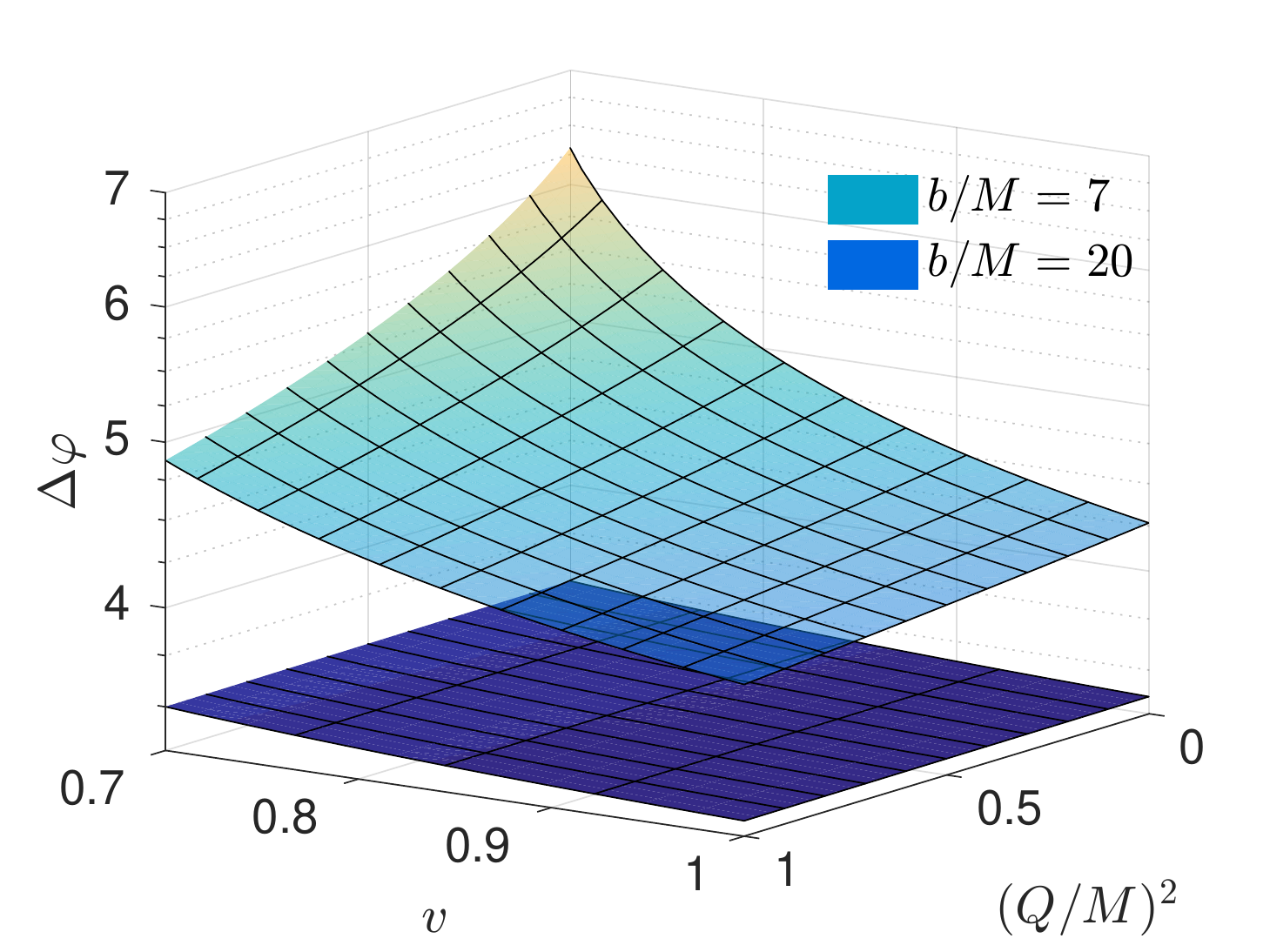}}\\
  \caption{The deflection angle. Fig. \ref{fig:deltaphih0}: $\Delta\varphi(h=0,b/M,v)$; Fig \ref{fig:deltaphibmanydiffv}: $\Delta\varphi(h,b/M=7,v)$ and  $\Delta\varphi(h,b/M=20,v)$. }
  \label{fig:deltaphi}
\end{figure}

Eq. \eqref{eq:deltaphiint} gives the total angular change for a particle travel through an unbounded orbit, but the net change of direction of the ray, is given by the modulated deflection angle $\Delta\varphi_{mod}$ defined as
\iea{
  \Delta\varphi_{mod}=\Delta\varphi-(2n+1)\pi. \label{eq:defdeltaphimod}
}
where $n$ is such that $-\pi<\Delta \varphi_{mod}\leq \pi$.

\section{Deflection in the weak and strong field limits} \label{sec:weakstrong}

\subsection{Weak field limit} \label{sec:deflectionweak}

The weak field limit happens for large $f=bv/M$.  Under this limit, the roots $\omega_i~ (i=1,2,3,4)$ for the quartic equation \eqref{eq:polyomega} can be solved perturbatively by using the asymptotic expansion method. To the order of $1/f^2$, the results are
\iea{
  \omega_{\substack{1\\2}}&=& \mp\frac{v}{f}+\frac{1}{f^2},\label{eq:rootv1finfexp12}\\
\omega_{\substack{3\\4}}&=& \frac{1}{1\pm\sqrt{1 -h^2}}-\frac{1}{f^2}\left (1\pm\frac{2-h^2}{2 \sqrt{1-h^2}}\right).\label{eq:rootv1finfexp34}
}
Substituting these approximations into deflection angle \eqref{eq:deltaphi}, to the order of $1/f^2$, we obtain
\iea{
  \Delta\varphi&=&\pi +\frac{2 }{f}\left (v+\frac{1}{v}\right)+\frac{\pi}{4f^2}\left[3\left (4+v^2\right)- h^2 (2+v^2)\right],\\
  &=&\pi+\frac{2 M }{b }\left (1+\frac{1}{v^2}\right)+\frac{ \pi }{4 b^2}\left[3M^2 \left (\frac{4}{v^2}+1\right)-Q^2 \left (\frac{2}{v^2}+1\right)\right].\label{eq:deltaphiv1finfexp}
}
The first two terms and first term in square bracket give the deflection angle of a particle moving in Schwarzschild spacetime with a large impact parameter $b$  (see Ref. \cite{Misner:1974qy} for the first two terms and Ref. \cite{Crisnejo:2018uyn} for a simple derivation of the result in the Schwarzschild case) and the last term in the square bracket gives the charge correction. Eq. \eqref{eq:deltaphiv1finfexp} agrees with the Eq. (12) of Ref. \cite{He:2016vxc} after setting its angular momentum $a$ to zero. Clearly, the deflection angle is reduced due to the increase of charge.

In order for the expansion \eqref{eq:deltaphiv1finfexp} to be valid, the higher order terms have to be smaller than the leading term so that the entire expansion can be convergent. This apparently requires $v>1/f$, i.e., $v^2>M/b$. Indeed, one can show that the limit $ (f,v)\to  (\infty,0)$ is not well defined for the deflection angle $\Delta \varphi$ in Eq. \eqref{eq:deltaphi}, in that if one chooses different path to approach this point in its domain space, $\Delta \varphi$ will end up with different values. Therefore, around this point, there can exist different expansions of $\Delta\varphi$  with different validity regions in the parameter space of $ (h,~f,~v)$. One has to be explicit and careful about where the expansion works.

\subsubsection{Deflection angles of relativistic particles}
Setting $v=1$ in Eq. \eqref{eq:deltaphiv1finfexp} we get the deflection angle of light
\iea{
  \left.\Delta\varphi\right|_{v=1}&=& \pi+\frac{4 M}{b}+\frac{3 \pi\left (5M^2-Q^2\right)}{4 b^2}+\mathcal{O}\left (\frac{1}{b^3}\right).\label{eq:deltaphiv1finfexpv1}
}
For a particle that moves relativistically, Eq. \eqref{eq:deltaphiv1finfexp} can be expanded around $v=1$ to produce the velocity correction to \eqref{eq:deltaphiv1finfexpv1}
\iea{
  \Delta\varphi&=&\left.\Delta\varphi\right|_{v=1}- (1-v)\left[-\frac{4M}{b}-\frac{\pi\left (6   M^2-  Q^2\right)}{b^2}\right]+\mathcal{O}\left (\frac{1}{b^3}\right). \label{eq:deltaphiv1finfexpv1vcorre}
}
From this, it is seen that for any given $b$ the deflection angle $\Delta\varphi$ increases when we decrease $v$. This is in agreement to what we see in Fig. \ref{fig:deltaphih0} for large $b$. Moreover, the increase of charge will cause a small decrease of the deflection angle at order $\mathcal{O} (1/b^2)$, which is in contrast to the order $\mathcal{O} (1/b)$ correction due to velocity but in align with the $b/M=20$ plot in Fig. \ref{fig:deltaphibmanydiffv}.

\subsubsection{Deflection angles of slower particles}

For smaller $v$ satisfying $v\gg1/f$, expansion \eqref{eq:deltaphiv1finfexp} can still be used. In particular, if $v$ is very small such that $1/f\ll v\ll 1$, then Eq. \eqref{eq:deltaphiv1finfexp} can be approximated by
\iea{
  \Delta\varphi&=& \pi+\frac{2 M }{bv^2}+\frac{ \pi (6 M^2 -Q^2)}{ 2b^2 v^2}+\mathcal{O}\left (\frac{1}{b^3}\right) .\label{eq:deltaphiv0finfexp1}
}
In this range of $v$, clearly, as $v$ decreases, the deflection angle increases at order $\mathcal{O} (1/v^2)$.
For even smaller $v$ such that $v\ll 1/f$ or $v^2\ll M/b$, the deflection angle can be obtained by follow a similar asymptotic expansion procedure. The result is
\iea{
  \Delta\varphi &=&2 \pi -\frac{2 b v^2}{M}+\frac{\pi   (6M^2-Q^2)}{b^2 v^2}+\mathcal{O}\left (\frac{1}{b^3}\right). \label{eq:deltaphiv0finfexp2}
}
For very small $v$ such that $v^2<  (M/b)^{3/2}$,  Eq. \eqref{eq:deltaphiv0finfexp2} is great than $2\pi$. This is in contrast to the Newtonian gravity where the unbounded orbits around a central body are always hyperbolics or parabolas and therefore the particle deflection angle will never exceed $2\pi$. In this sense therefore $\Delta\varphi$ exceeding $2\pi$ is purely a general relativistic effect. Although the correction to the deflection angle from charge $Q$ in Eqs. \eqref{eq:deltaphiv0finfexp1} and \eqref{eq:deltaphiv0finfexp2} are also at order $\mathcal{O} (1/b^2)$, however, unlike in Eq. \eqref{eq:deltaphiv1finfexpv1vcorre}, there is no suppression but a boost due to small $v$ by order $\mathcal{O} (1/v^2)$. Around the upper bound of the validity of $v$, i.e., $v\approx\mathcal{O} (1/\sqrt{b})$, these corrections can also be comparable to the leading term. This can be seen from the deviation of the plots of  Eqs. \eqref{eq:deltaphiv0finfexp1} and \eqref{eq:deltaphiv0finfexp2} from that of the exact result in Fig. \ref{fig:deltaphiv0exphhalf}.

\subsection{Strong field limit}\label{sec:strong}

\subsubsection{Deflection angles of relativistic particles}

The strong field limit happened when \mb approaches $b_c (h,v)$ in Eq. \eqref{eq:defbc}. For relativistic particles, the roots can be given approximately as a series of $ (1-v)$ and the small deviation $\delta (h,v)$ of $f (h,v)$ from $f_c (h,v)$
\be
\delta (h,v)=f-f_c (h,v) \label{deltahvdef}
\ee
as the following
\iea{
	\omega_{\substack{1\\4}}&=&\frac{x}{x+2} \left[-\frac{2}{ x^2+2}-\frac{   (x+2)^3 (v-1)}{2 (x+1) \left (x^2+2\right)^2}\pm\frac{ \sqrt{2}  x  (x+2)^3\delta (h,v) }{4 (x+1)^2 \left (x^2+2\right)^{5/2}}\right.\nonumber\\
	&&\left.\mp\frac{\sqrt{2}  x \left (5 x^4-12 x^3-52 x^2-48 x-16\right)  (x+2)^3  (v-1)\delta (h,v) }{32 (x+1)^4 \left (x^2+2\right)^{7/2}}\right], \label{eq:roots14}\\
	\omega_{\substack{2\\3}}&=& \frac{2}{ x^2+2}\mp\frac{ 2^{3/4} x^{3/2} \sqrt{\delta (h,v)}}{ (x^2-1)\left (x^2+2\right)^{7/4}}+\frac{  x^4 (v-1)}{2 (x^6+3 x^4-4)}-\frac{ \sqrt{2}  \left (x^2-2\right) x^3\delta (h,v)}{2\left (x^2-1\right)^2 \left (x^2+2\right)^{5/2}}\nonumber \\
	&&\pm\frac{2^{3/4} \left (4 x^4-5 x^2-2\right) x^{7/2} (v-1)\sqrt{\delta (h,v) }}{8\left (x^2-1\right)^{5/2} \left (x^2+2\right)^{11/4} }-\frac{\sqrt{2}  \left (3 x^6-11 x^4+22 x^2-20\right) x^5  (v-1)\delta (h,v) }{8\left (x^2-1\right)^4 \left (x^2+2\right)^{7/2}} , \label{eq:roots23}
  }
where
\iea{
  x=\sqrt{\sqrt{9-8 h^2}+1}. \label{eq:xdef}
}
 Despite their long form,  the calculation of these four expansions is straight forward and will be illustrated in appendix \ref{sec:rootsv1pert}. With these expansions of roots, the deflection angle \eqref{eq:deltaphi} can be easily expanded as
\iea{
  \Delta\varphi (h,f,v\to 1)&=& \Delta\varphi [h, f=f_c (h,v)+\delta (h,v), v=1]-  (1-v) a_1 (h,v)+\mathcal{O}\left[ (1-v)^2\right] \label{eq:deltaphiv1fcexp}
}
with the leading term
\bea
\Delta\varphi [h, f=f_c (h,v)+\delta (h,v),v=1]&=& \frac{1}{2 \sqrt{2}}\sqrt{\frac{x^2+2}{x^2-1}} \left\{-2 \ln [\delta (h,v) ]+\ln \left[\frac{2^{13} \left (x^2-1\right)^6 \left (x^2+2\right)^3}{x^{10} \left (\sqrt{x^2-1}+x\right)^4}\right]\right\}+O[\delta (h,v)],\label{eq:deltaphiv1fcexpv1} \nonumber \\
\eea
and $a_1$ the coefficient of the velocity expansion
\bea
a_1 (h,v)&=&-\frac{1}{4 \sqrt{2}\left (x^2-1\right)^{2}  (x^2+2)}\left\{ \frac{\sqrt{2}}{2}\left (2 x^4-x^2-4\right)x^2 \Delta\varphi [h, f=f_c (h,v)+\delta (h,v), v=1]-\sqrt{\frac{x^2+2}{x^2-1}}a_1' \right\},\label{eq:defa1}\nonumber\\
\eea
where
\bea
a'_1&=&\left (11 x^4-26 x^2+22\right) x^2+\left (8-5 x^2\right)x^3\sqrt{x^2-1} -16. \label{eq:defa1p}
\eea

Eq. \eqref{eq:deltaphiv1fcexp} gives the velocity correction of the deflection angle of massive particles respect to light.
However, because it is only a partial expansion of velocity and there is still velocity dependency in $\delta (h,v)$ and $a_1 (h,v)$, it is not easy to extract the full effect of velocity.  Furthermore, recall $\delta (h,v)=f-f_c (h,v)$ and $f=bv/M$ we find that $f$ is also a function of $v$ if we fix $b$. We then define
 \iea{
   \delta_b (h,v)&=& \frac{b-b_c (h,v)}{M},\label{eq:defdeltab}
 }
 and using the relation \eqref{eq:defbc} between $b_c (h,v)$ an $f_c (h,v)$, we get a simple relation $\delta (h,v)=\delta_b (h,v)v$. Substituting this back into Eq. \eqref{eq:deltaphiv1fcexp} and further expanding it respect to $ (1-v)$ we find
\iea{
\Delta\varphi (h,b,v\to 1)
&=& \Delta\varphi [h, b=b_c (h,1)+\delta_b (h,1), v=1]\nonumber\\
&&-  (1-v) \left[- \frac{ (x^2+2) (x^2+4)}{4x\sqrt{x^2-1}\delta_b (h,1)} + a_1 (h,1)-\sqrt{\frac{x^2+2}{2 (x^2-1)}} +\mathcal{O}\left (\delta_b\right)\right] +\mathcal{O}\left[ (1-v)^2\right], \label{eq:deltaphiv1fcexpvcorrecbcon}
}
where
\be
\Delta\varphi [h, b=b_c (h,1)+\delta_b (h,1), v=1]=\frac{1}{2 \sqrt{2}}\sqrt{\frac{x^2+2}{x^2-1}} \left\{-2 \ln [\delta_b (h,v) ]+\ln \left[\frac{2^{13} \left (x^2-1\right)^6 \left (x^2+2\right)^3}{x^{10} \left (\sqrt{x^2-1}+x\right)^4}\right]\right\}+O[\delta_b (h,v)],\label{eq:deltaphiv1fcexpv1b}
\ee
and
\be
\delta_b (h,1)=\frac{b-b_c (h,v)}{M}=\frac{b}{M}-\frac{ (x^2+2)^{3/2}}{\sqrt{2}x}.
\ee
can be obtained from Eq. \eqref{eq:deltaphiv1fcexpv1} and \eqref{eq:defdeltab} respectively by letting $v=1$.

The first term  $\Delta\varphi[h,b=b_c (h,1)+\delta_b (h,1),v=1]$ gives the deflection of light in strong field limit. Unlike	Eq.  (61) obtained in  \cite{Bozza:2002zj}, the logarithm term in Eq. \eqref{eq:deltaphiv1fcexpv1b} is accurate to any order of $h$. Furthermore, our result \eqref{eq:deltaphiv1fcexpv1b} is exactly the same as the one obtained by Tsukamoto and Gong  (\cite{Tsukamoto:2016oca}, Eq.  (2.43)). For the velocity correction term in Eq. \eqref{eq:deltaphiv1fcexpvcorrecbcon}, we can see that its coefficient is large if $\delta_b$ is small enough. Since this coefficient is negative, the deflection angle will increase as $v$ deviates from $1$, which agrees with what we observed in Fig. \ref{fig:deltaphi}.
This $1/\delta_b$ term indeed comes from the velocity expansion of term $\ln (f-f_c (h,v))=\ln[ (b-b_c (h,v)v/M)]$ in Eq. \eqref{eq:deltaphiv1fcexp}. Therefore, in order for the entire expansion \eqref{eq:deltaphiv1fcexpvcorrecbcon} to be valid, we have to demand $\mathcal{O}[ (1-v)/\delta_b]<\mathcal{O}\left[-\ln  (\delta_b)\right]$. This implies that \mathv cannot deviate from $1$ too much, i.e., $v>\mathcal{O}\left[1+\delta_b\ln (\delta_b)\right]$.

To find out the influence of charge on the deflection angle in this case, we expand \eqref{eq:deltaphiv1fcexpvcorrecbcon} near $h=0$ to the order of $h^2$
\iea{
  \Delta\varphi (h\to0,b=b_c (0,1)+\delta_b,v\to1)&=&-\ln \left  (\delta_b\right)+\ln\left[648 (7\sqrt{3}-12)\right] -\frac{h^2}{18}\left (\frac{9\sqrt{3}}{\delta_b}+ \xi\right) -\frac{1}{9}  (1-v)\left[-\frac{18\sqrt{3}}{\delta_b} \right.  \nonumber\\
  && \left.+\xi-9-\frac{h^2}{18} \left (-\frac{486}{\delta^2_b}-\frac{18\sqrt{3}}{\delta_b}-\xi+15 \right)\right]+\mathcal{O}\left\{ (1-v)^2,h^4\right\},\label{eq:deltaphiv1fcexpvcorrech0exp}
}
with $\delta_b=b-b_c (0,1)=b-3\sqrt{3}$ and
\iea{
  \xi&=& 2 \ln \left (\delta_b\right) + \ln \left (\frac{97+56\sqrt{3}}{2^6\cdot3^9}\right)-4 \sqrt{3}+15.
}
We can see the leading term in the coefficient of $h^2$ is negative and thus the deflection decreases as the charge increases in strong field limit for given velocity $v$ and $f$ or equivalently the impact parameter $b=fM/v$. This is in agreement to what we observed in Fig. \ref{fig:deltaphibmanydiffv}.
Even though the charge correction is of order $h^2$, the $\ln\delta$ term in the coefficient might cause a comparable correction to the leading order. For exactly $v=1$ and $h=0$ we recover the deflection angle of Schwarzschild spacetime in the strong field limit \cite{Jia:2015zon}
\iea{
  \Delta\varphi (h=0,b=b_c (0,1)+\delta_b,v=1)&=& -\ln \left  (\delta_b\right)+\ln\left[648 (7\sqrt{3}-12)\right].\label{eq:deltaphiv1h0}
}

On the other hand, for near extremal RN black holes, the deflection angle can be expanded around $h=1$ to the order $ (1-h)$ as
\iea{
  && \Delta\varphi (h\to1,b= b_c (1,1)+\delta_b,v\to1) \nonumber \\
  &=&-\sqrt{2}\ln (\delta_b)+\sqrt{2}\ln\left[128 (3-2\sqrt{2})\right] - (1-h) \left\{-\frac{4 \sqrt{2}}{\delta_b }-3 \sqrt{2} \ln  (\delta_b )+3\sqrt{2}\ln\left[128\left (3-2\sqrt{2}\right)^2\right]\right.\nonumber \\
  &&\left.-17 \sqrt{2}+8+6 \sqrt{2} \sinh ^{-1} (1)\right\}- (1-v) \left\{-\frac{3\sqrt{2}}{\delta_b }+\frac{\ln (\delta_b)-\ln\left[128 (3-2\sqrt{2})\right]-2  (1+\sqrt{2})}{4 \sqrt{2}}\right.\nonumber \\
  &&\left.- (1-h) \left[\frac{12 \sqrt{2}}{\delta_b ^2}-\frac{15\sqrt{2}}{2 \delta_b }+\frac{-11 \ln  (\delta_b)+99\ln  (2)-44 \ln \left (2+\sqrt{2}\right)+14 \sqrt{2}-81+22\sinh ^{-1} (1)}{4 \sqrt{2}}\right]\right\}\nonumber \\
  &&+\mathcal{O}\left\{ (1-v)^2, (1-h)^2\right\}, \label{eq:deltaphiv1fcexpvcorrech1exp}
}
with $\delta_b=b-b_c (1,1)=b-4$. The leading two terms are the deflection of light ray for extremal RN black hole
\iea{
  \Delta\varphi (h=1,b= b_c (1,1)+\delta_b,v=1)&=&-\sqrt{2}\ln (\delta_b)+\sqrt{2}\ln[128 (3-2\sqrt{2})].\label{eq:deltaphiv1h1}
}
The leading term in the coefficient of $ (1-h)$ is positive, therefore $\Delta\varphi$ increases for charge deviates from $1$, again consistent with our observation in Fig. \ref{fig:deltaphibmanydiffv}.

Comparing Eq. \eqref{eq:deltaphiv1h1} with \eqref{eq:deltaphiv1h0} it is seen that the light deflection of extremal RN spacetime, namely, $\Delta\varphi (h=1,b=b_c (1,1)+\delta_b,v=1)$, is larger than that of Schwarzschild spacetime, namely, $\Delta\varphi (h=0,b=b_c (0,1)+\delta_b,v=1)$, for the same small $\delta_b$. However, one thing to be noted is that the two $\delta_b$'s in Eqs. \eqref{eq:deltaphiv1h0} and \eqref{eq:deltaphiv1h1} are defined respectively respect to different values of $b_c$, i.e., $b_c (h=1,v=1)=4M$ and $b_c (h=0,v=1)=3\sqrt{3}M$ . If one fixed suitable impact parameters $b$ and consequently $f$ were used, light deflection in extremal RN spacetime will be always smaller than that in Schwarzschild spacetime.

\begin{figure}[htp]
  \centering
  \subfloat[]{\label{fig:deltaphiv1exphhalf}\includegraphics[width=0.45\textwidth]{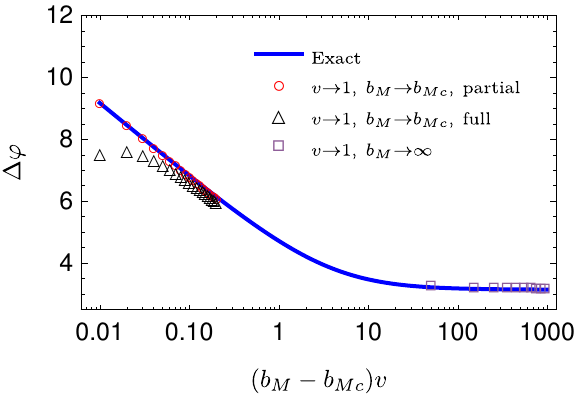}}~
  \subfloat[]{\label{fig:deltaphiv1expfceh0}\includegraphics[width=0.45\textwidth]{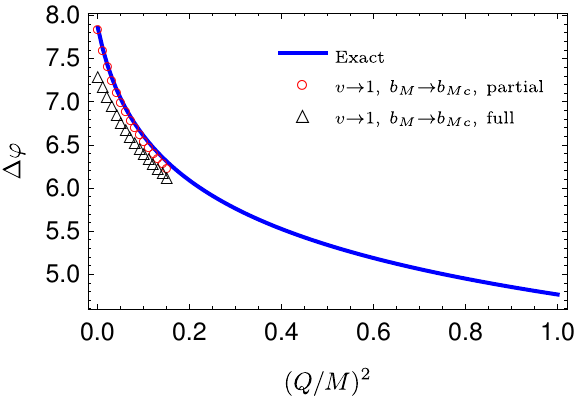}}\\
  \caption{The deflection angle and their expansions for relativistic particles. Fig. \ref{fig:deltaphiv1exphhalf}: $\Delta\varphi(h=1/2,b/M,v=0.99)$;  Fig. \ref{fig:deltaphiv1expfceh0}: $\Delta\varphi(h,b/M=0.56,v=0.99)$. In both plots the blue solid curve  are drawn from the exact formula \eqref{eq:deltaphi}. The red circle and  black triangle symbols are calculated from expansions \eqref{eq:deltaphiv1fcexp} and \eqref{eq:deltaphiv1fcexpvcorrecbcon} respectively. In Fig. \ref{fig:deltaphiv1exphhalf} the violet square symbol is obtained from the large $b_M$ expansion \eqref{eq:deltaphiv1finfexp}, which is absent in Fig. \ref{fig:deltaphiv1expfceh0} because for $b/M=5.26$ the field is still strong that can't be handled through weak field limit.}
  \label{fig:deltaphiv1exp}
\end{figure}

In  Fig. \ref{fig:deltaphiv1exphhalf} we plot the deflection angle as a function of $\delta_b (h,v)=[b-b_c (h,v)]v/M$ for relativistic particles with $v=0.99$ and  $h=1/2$ using exact deflection angle \eqref{eq:deltaphi} and expansions  \eqref{eq:deltaphiv1finfexp}, \eqref{eq:deltaphiv1fcexp} and \eqref{eq:deltaphiv1fcexpvcorrecbcon}. At large $b$, it is seen that expansion \eqref{eq:deltaphiv1finfexp}  (the square symbols) fits the exact deflection angle  (the solid line) very well. At smaller $b$, expansion \eqref{eq:deltaphiv1fcexp}  (the circle symbols) which does not fully expand its $v$ dependence fits the exact result well too. While for expansion \eqref{eq:deltaphiv1fcexpvcorrecbcon}  (the triangle symbols), it is seen that as $b$ decreases towards $b_c$, which is equivalent to $\mathcal{O}\left[ (1-v)/\delta_b\right]$ approaching $\mathcal{O}[-\ln (v)]$  (roughly at $\delta_b (h,v)\approx 1/200$), the deviation of this expansion from the exact deflection angle becomes apparent  and will eventually be comparable to the exact deflection angle.
Fig. \ref{fig:deltaphiv1expfceh0} shows the effect of charge to the deflection angle. Again, the expansion  \eqref{eq:deltaphiv1fcexpvcorrecbcon} departs from the exact angle. This is understandable because the decrease of $Q$ will decrease $\delta_b=b-b_c (h,v)$ for fixed $b$ such that the condition $v>\mathcal{O}[1+\delta_b\ln (\delta_b)]$ becomes more violated.

\subsubsection{Deflection angles of non-relativistic particles}

Unlike expansion \eqref{eq:deltaphiv1finfexp} in the weak field limit, which works for both relativistic and non-relativistic $v$, expansion \eqref{eq:deltaphiv1fcexp} is only valid when $ (1-v)$ is small. Therefore, we have to carry out the small velocity expansion separately for the strong field limit. Using the exact deflection angle \eqref{eq:deltaphi}, the result is found to be
\iea{
  \Delta\varphi (h,f=f_c (h,0)+\delta (h,0),v\to0)&=& \frac{-2 \ln \left[\frac{2 \omega_{2h} \omega _4}{\omega_{c0} \left (\omega_{c0}-\omega _4\right)}\right]-\ln [\delta (h,0) ]+8 \ln  (2)}{h \sqrt{\omega_{c0} \left (\omega _4-\omega_{c0}\right)}}-\frac{2 v \sqrt{1-h^2 \omega_{c0}}}{\sqrt{\omega_{c0} \left (2 h^2 \omega_{c0}^2-3 \omega_{c0}+1\right)}}, \label{eq:deltaphifcv0expomegac}
}
where $\delta (h,0)=f-f_c (h,0)$, $\omega_{c0}=M/r_c (h,v=0)$ is the reciprocal of the radius of particle sphere \eqref{eq:rcv0}, $\omega_4$ is the fourth roots of Eq. \eqref{eq:polyomega} at $v=0$
\iea{
  \omega_4&=&  \frac{-20 h^6 \omega_{c0}^3+6 h^4 \omega_{c0} \left (5 \omega_{c0}-3\right)+8 h^2 \left (\omega_{c0}+1\right)-8}{\left\{\left[\omega_{c0} \left (6 h^4 \omega_{c0}^2-9 h^2 \omega_{c0}+4\right)+3\right]h^2 -4\right\}h^2},
}
and
\iea{
  \omega_{2h}&=& -\frac{\sqrt{2} \omega_{c0}^{5/4} \sqrt{-\sqrt{h^2 \omega_{c0}-1} \sqrt{\omega_{c0} \left (3-2 h^2 \omega_{c0}\right)-1} \left[\omega_{c0} \left (h^2 \omega_{c0}-2\right)+1\right]}}{\sqrt{\omega_{c0} \left[h^2 \omega_{c0} \left (4 h^2 \omega_{c0}-9\right)+6\right]-1}}.
}

Again, we are interested in the small charge and extremal RN limits of this deflection angle. For $h\to0$ we have
\iea{
  &&\Delta\varphi (h\to0,f=4+\delta (0,0),v\to0)\nonumber \\
  =&& -\sqrt{2} \ln \left [\delta (0,0) \right]+7\sqrt{2}\ln (2) +\left\{\frac{7 \ln  (2)}{8 \sqrt{2}}-\frac{1}{\sqrt{2} \delta  (0,0) }-\frac{\ln  [\delta (0,0) ]}{8 \sqrt{2}}-\frac{3}{4 \sqrt{2}}\right\}h^2 + (-8+h^2) v +\mathcal{O}\{h^3,v^2\}, \label{eq:deltaphifcv0expomegach0expfc4}
}
where $\delta (0,0)= f-f_c (0,0) = f-4$. Setting  exactly $h=0$, we obtain the strong field deflection angle of non-relativistic particles in Schwarzschild spacetime
\iea{
  \Delta\varphi (h=0,f=4+\delta (0,0),v\to0)&=& -\sqrt{2}\ln (\delta (0,0))+7\sqrt{2}\ln (2)-8v+\mathcal{O} (v^2).
}
Comparing to Eq. \eqref{eq:deltaphiv1h1}, we can see that the deflection angle of a non-relativistic particle in Schwarzschild spacetime is close to that of light in the extremal RN spacetime. Note that in these two cases the $f$'s are accidently equal because of the relation $f_c (1,1)=f_c (0,0)=4$.

For $h\to 1$, the expansion \eqref{eq:deltaphifcv0expomegac} becomes
\iea{
  &&\Delta\varphi (h\to1,f=f_c (1,0)+\delta,v\to0)\nonumber \\
  =&&-\frac{1}{20} \sqrt[4]{5} \left (5+3 \sqrt{5}\right) \left [2\ln (\delta)-3 \ln  (20)+\sinh ^{-1} (2)\right]- (1-h)\left\{-\frac{\sqrt[4]{5}\sqrt{2+\sqrt{5}} \left (5+3 \sqrt{5}\right)}{10 \delta }\right.\nonumber \\
  &&\left.+\frac{\sqrt[4]{5}}{100}  \left (45+23 \sqrt{5}\right) \left [-2\ln  (\delta )+3\ln (20)-\frac{133-17\sqrt{5}}{62}-\sinh ^{-1} (2)\right]~\right\} \nonumber \\
  && + v\left[-\sqrt{22+10 \sqrt{5}}-2\sqrt{2+\sqrt{5}} (1-h)\right]+\mathcal{O}[ (1-h)^2,v^2],\label{eq:deltaphifcv0expomegach1expfch1}
}
where $\delta (1,0)=f-f_c (0,1)=f-\sqrt{\left (11+5\sqrt{5}\right)/2}$. Setting exactly $h=1$, we get the deflection angle for non-relativistic particles in the extremal RN spacetime
\iea{
  &&\Delta\varphi (h=1,f=f_c (1,0)+\delta,v\to0) \nonumber \\
  =&&-\frac{1}{20} \sqrt[4]{5} \left (5+3 \sqrt{5}\right) \left[2\ln (\delta)-3 \ln  (20)+\sinh ^{-1} (2)\right]- \sqrt{22+10 \sqrt{5}}v+\mathcal{O} (v^2). \label{eq:deltaphifcv0expomegach1expfch1h1}
}

\begin{figure}[htp]
  \centering
  \subfloat[]{\label{fig:deltaphiv0exphhalf}\includegraphics[width=0.45\textwidth]{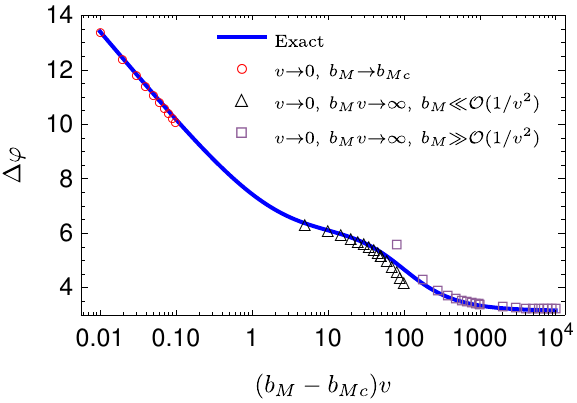}}~
  \subfloat[]{\label{fig:deltaphiv0expfceh0}\includegraphics[width=0.45\textwidth]{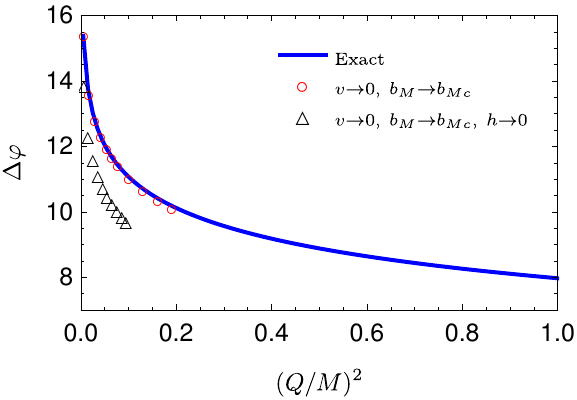}}\\
  \caption{The deflection angle and their expansions for non-relativistic particles. Fig. \ref{fig:deltaphiv0exphhalf}: the deflection angle with $h=1/2$ and $v=0.01$. Fig. \ref{fig:deltaphiv0expfceh0}: the deflection angle with $b/M=400$ and $v=0.01$. Blue solid curve: the exact formula \eqref{eq:deltaphi};  Red circle symbol: small $ (b_M-b_{Mc})v$ expansion \eqref{eq:deltaphifcv0expomegac}; Black triangle: large $b_M$ expansions \eqref{eq:deltaphiv0finfexp1};  Violet square symbol: expansion \eqref{eq:deltaphiv0finfexp2}. Fig. \ref{fig:deltaphiv0expfceh0}: Black triangle symbol: small $Q$ expansion \eqref{eq:deltaphifcv0expomegach0expfc4}. No large $Q$ expansion \eqref{eq:deltaphifcv0expomegach1expfch1} is plotted in Fig. \ref{fig:deltaphiv0expfceh0} because $b/M=400$ is too large to be approximated by strong field limit for $Q$ approaching $1$.}
  \label{fig:deltaphiv0exp}
\end{figure}

In Fig. \ref{fig:deltaphiv0exp}, we plot the deflection angle for non-relativistic particles with $v=0.01$. In Fig. \ref{fig:deltaphiv0exphhalf}, $h$ is fixed at $1/2$. The blue solid curve is drawn from the exact formula \eqref{eq:deltaphi}. The red circle  symbol is calculated from the small $\delta= (b_M-b_{Mc})v$ expansion \eqref{eq:deltaphifcv0expomegac}, and  black triangle and  violet square symbols are obtained from the large $b_M$ expansions \eqref{eq:deltaphiv0finfexp2} and \eqref{eq:deltaphiv0finfexp1} respectively. It is seen that all three expansions work quite well in their valid regions we discussed previously. However, outside these regions, e.g., when $ (b_M-b_{Mc})v\approx100$, the valid condition about expansion \eqref{eq:deltaphiv0finfexp2}  ($b_M\ll\mathcal{O} (1/v^2)$) that of expansion \eqref{eq:deltaphiv0finfexp1}  ($b_M\gg\mathcal{O} (1/v^2)$) are both violated. One can see from Fig. \ref{fig:deltaphiv0exphhalf} that at this point, both their numerical values deviation from the exact value noticeably.
Moreover, from the red circle in Fig. \ref{fig:deltaphiv0exphhalf} we can see that $\Delta\varphi$ diverges logarithmically indeed.  Fig. \ref{fig:deltaphiv0expfceh0} shows the effect of charge to the deflection angle of non-relativistic rays with $v=0.01$ and $b/M=400$. If this particle ray is in spacetime with $Q=0$, then its critical impact parameter $b_c/M\approx399.3$, which is very close to its actually $b=400M$. Therefore, the ray will loop around the black hole many times, resulting in a large deflection angle. However, if $Q$ approached $1$, we have $b_c/M\approx333.0$, which is quite away from $400$. Consequently, the particle ray will experience a much smaller deflection angle compared to the rays with same $b$ but in spacetimes with smaller $Q$.

\section{Apparent angles} \label{sec:apparentangle}

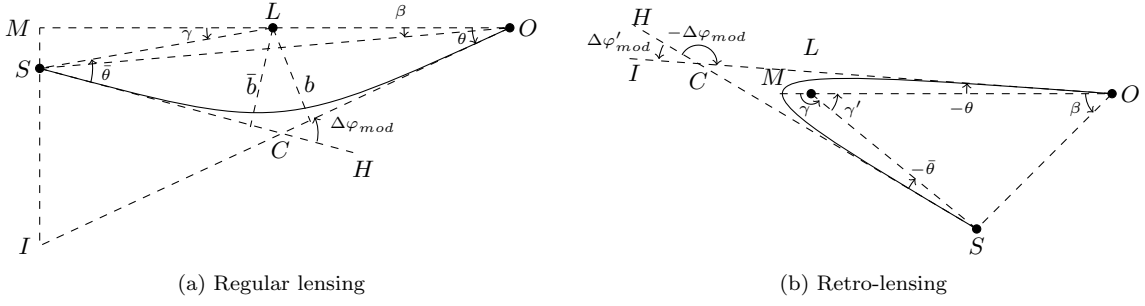
\begin{figure}[htp]
  \subfloat[Regular lensing]{\label{fig:lensreg}
\begin{tikzpicture}[dot/.style={circle,inner sep=1pt,fill,label={#1},name=#1},
  extended line/.style={shorten >=-#1,shorten <=-#1},
 extended line/.default=1cm,
  one end extended/.style={shorten >=-#1},
 one end extended/.default=1cm,
  ]
	\coordinate (L) at (0,0);
	\coordinate (O) at (3.15,0);
	\coordinate (S) at (-3.1,-0.54);
	\coordinate (M) at (-3.1,0);
	\coordinate (C) at (0.10,-1.4);
	\coordinate (I) at (-3.1,-2.9);
	\coordinate (H) at (1.2,-1.65);

  \draw[dashed] (L)--(S)--(O)--(M);
  \draw (S) .. controls (C) ..(O);
  \draw[one end extended=1cm,dashed] (S) --(C);
  \draw[dashed] (O) --(I)--(M);
  \draw[fill] (L) circle[radius=0.055];
  \node[above] at  (L) {$L$};
  \draw[fill] (S) circle[radius=0.055];
  \node[left] at  (S) {$S$};
  \draw[fill] (O) circle[radius=0.055];
  \node[right] at  (O) {$O$};
  \node[below] at  (C) {$C$};
  \node[left] at  (M) {$M$};
  \node[below] at  (H) {$H$};
  \node[left] at  (I) {$I$};

	\pic[draw,->,"$\scriptstyle\theta$",angle eccentricity=1.3] {angle=L--O--C};
	\pic[draw,->,angle eccentricity=1.2,angle radius=40] {angle=L--O--S};
	\node[above] at (1.7,0) {$\scriptstyle\beta$};
	\pic[draw,->,angle eccentricity=1.3,angle radius=20] {angle=C--S--L};
	\node[right] at (-2.4,-0.6) {$\scriptstyle\bar{\theta}$};
	\pic[draw,->,"$\scriptstyle\Delta\varphi_{mod}$",angle eccentricity=2.2] {angle=H--C--O};
	\pic[draw,->,"$\scriptstyle\gamma$",angle eccentricity=1.3,angle radius=25] {angle=M--L--S};
	\draw[dashed] (L)--(-0.3,-1.32) coordinate (bbper);
	\node[left] at (-0.1,-0.75) {$\bar{b}$};
	\draw[dashed] (L)--(0.5,-1.25);
	\node[right] at (0.3,-0.75) {$b$};
\end{tikzpicture}
}~
\subfloat[Retro-lensing]{\label{fig:lensretro}
\begin{tikzpicture}[dot/.style={circle,inner sep=1pt,fill,label={#1},name=#1},
  extended line/.style={shorten >=-#1,shorten <=-#1},
 extended line/.default=1cm,
  one end extended/.style={shorten >=-#1},
 one end extended/.default=1cm,
  ]

  \coordinate (L) at (0,0);
  \coordinate (O) at (4,0);
  \coordinate (S) at (2.2,-1.8);
  \coordinate (C) at (-1.5,0.4);
  \coordinate (M) at (-0.5,0);
  \coordinate (H) at (-2.25,0.8);
  \coordinate (I) at (-2.35,0.50);

  \draw[one end extended=0.5cm,dashed] (L)--(S)--(O)--(L);
  \draw (S) .. controls (C) ..(O);
  \draw[one end extended=1cm,dashed] (S) --(C);
  \draw[one end extended=1cm,dashed] (O) --(C);
  \draw[fill] (L) circle[radius=0.055];
  \node[above] at  (0,0.4) {$L$};
  \draw[fill] (S) circle[radius=0.055];
  \node[below] at  (S) {$S$};
  \draw[fill] (O) circle[radius=0.055];
  \node[right] at  (O) {$O$};
  \node[below] at  (C) {$C$};
  \node[above] at  (M) {$M$};
  \node[above] at  (H) {$H$};
  \node[below] at  (I) {$I$};

  \pic[draw,<-,angle eccentricity=1.3,angle radius=55] {angle=C--O--L};
  \node[below] at (2,0) {$\scriptstyle-\theta$};
  \pic[draw,->,"$\scriptstyle\beta$",angle eccentricity=1.5,angle radius=10] {angle=L--O--S};
  \pic[draw,<-,angle eccentricity=1.2,angle radius=30] {angle=L--S--C};
  \node[right] at (1.2,-1) {$\scriptstyle-\bar{\theta}$};
  \pic[draw,<-,"$\scriptstyle-\Delta\varphi_{mod}$",angle eccentricity=1.7,angle radius=7] {angle=O--C--H};
  \pic[draw,->,angle eccentricity=2.2,angle radius=15] {angle=H--C--I};
  \node[left] at (-2,0.7) {$\scriptstyle\Delta\varphi^\prime_{mod}$};
  \pic[draw,->,"$\scriptstyle\gamma$",angle eccentricity=1.8,angle radius=4] {angle=M--L--S};
  \pic[draw,->,"$\scriptstyle\gamma^\prime$",angle eccentricity=1.7,angle radius=10] {angle=S--L--O};
\end{tikzpicture}
}
\caption{Lensing configuration: Fig. \ref{fig:lensreg} shows the configuration of regular lensing, while Fig. \ref{fig:lensretro} shows the configuration of retro-lensing. In both figures,  $O,~L,~S$ and $I$ stand for the observer, lens, source and image respectively, and the optics axis is chosen as the line that join $O$ and the lens $L$. The observer $O$ and the source $S$ are sitting in the flat spacetime region. $\theta$ is the image position and $\beta$ is the source position if the spacetime is flat. $\Delta\varphi_{mod}$ is the modulated deflection angle. The angle is positive if it follows the direction indicates by the arrow and circles counterclockwise, therefore in Fig. \ref{fig:lensretro} $\Delta\varphi_{mod},~\theta$ and $\bar{\theta}$ are companied by minus signs. Further in Fig. \ref{fig:lensretro} we also show angles $\Delta\varphi_{mod}^\prime$ and $\gamma'$ defined by Eq. \eqref{eq:defdeltaphimodprime}. Note that in Fig. \ref{fig:lensretro} we have $d_{\rm ls}<d_{\rm ol}$, therefore $\beta'=\beta$ according to Eq. \eqref{eq:defdeltaphimodprime}.}
  \label{fig:lenscon}
\end{figure}

Fig. \ref{fig:lensreg} illustrates a typical geometric configuration of GL. A particle starting from source $S$ with impact parameter $\bar{b}$ will be deflected by the lens $L$ resulting in a modulated deflection angle $\Delta\varphi_{mod}$ and received by the observer $O$. The angular position of the source is denoted by $\beta$ and that of the image by $\theta$ and termed as {\it apparent angle}. Note we only need to study the case $\beta\geq0$ because the situation with $\beta<0$ is mirror symmetric to the corresponding $\beta>0$ case.
$O$ and $S$ are assumed far away from the lens object $L$, and thus the angles can be related by Euclidean geometry. Denoting angles $\angle MLS$ and $\angle CSL$ by $\gamma$ and $\bar{\theta}$ respectively,  we have the Ohanian lensing equation \cite{Ohanian:1987tb}
\iea{
  \gamma&=& \theta+\bar{\theta}-\Delta\varphi_{mod}.\label{eq:Ohanian}
}
Denoting the distances between $L$ and $S$, and $O$ and $L$ as $d_{\rm ls}$ and $d_{\rm ol}$ respectively, then  angles $\gamma$ and  $\beta$ are related through triangle $\triangle OSL$ by the sine law
\iea{
  \frac{d_{\rm ol}}{\sin(\gamma-\beta)}&=&  \frac{d_{\rm ls}}{\sin\beta}.\label{eq:betagamma}
}
Angles  $\bar{\theta}$ and $\theta$ are related  to the impact parameter $\bar{b}$ and $b$ respectively and due to the conservation of angular momentum
\iea{
  d_{\rm ls}\sin\bar{\theta}=b&=& d_{\rm ol}\sin\theta.\label{eq:thetathetabar}
}

If the source and observer are situated at two opposite sides of the lens, and $\gamma,~\beta,~\bar{\theta}$ and $\theta$ are all small angles, we say that the lensing is regular. In this case Eqs. \eqref{eq:betagamma} and \eqref{eq:thetathetabar}  reduce to
\iea{
  \gamma&=& \frac{d_{\rm ls}+d_{\rm ol}}{d_{\rm ls}}\beta,\quad\bar{\theta}= \frac{d_{\rm ol}}{d_{\rm ls}}\theta. \label{eq:gammabetasmall}
}
Substituting into Eq. \eqref{eq:Ohanian} we get the regular lensing equation
\iea{
  \beta&=& \theta-\frac{d_{\rm ls}}{d_{\rm ls}+d_{\rm ol}}\Delta\varphi_{mod}. \label{eq:lensingnormalsmall}
}
Note that there exists different approaches using other relations of the lensing geometry for the derivation of this equation, e.g., the sine law of triangle $\triangle OSC$ was used in Ref. \cite{Virbhadra:1998dy}. However to the lowest order under the assumptions that $\beta,~\theta,~\Delta\phi_{mod}$ are small and $d_{ol}, ~d_{ls}$ are larger, all approaches should lead to equation \eqref{eq:lensingnormalsmall}.
If the source and observer are situated at the same side of the lens and the relevant angles are small, then retro-lensing happens. In this case, $\gamma$ and $\Delta\varphi_{mod}$ take values around $\pi$ and $-\pi$ respectively, while $\beta$ takes value around $0$ or $\pi$ depending on whether $d_{\rm ls}<d_{\rm ol}$ or $d_{\rm ls}>d_{\rm ol}$ (we consider the case $|d_{\rm ls}-d_{\rm ol}|\gg1$ for simplicity). Introducing small angles $\gamma',~\Delta\varphi_{mod}',~\beta'$ as
\iea{
  \gamma'&=& \gamma-\pi,
  ~\Delta\varphi_{mod}'= \Delta\varphi_{mod}+\pi,
  \beta'=\beta~\text{if}~d_{\rm ls}<d_{\rm ol}~\text{or}~\beta'=\beta-\pi~\text{if}~d_{\rm ls}>d_{\rm ol}, \label{eq:defdeltaphimodprime}
}
Then the Ohanian Eq. \eqref{eq:Ohanian} is unchanged in terms of $\gamma'$ and $\Delta\varphi_{mod}'$
\iea{
  \gamma'=\theta+\bar{\theta}-\Delta\varphi_{mod}'. \label{eq:Ohanianp}
}
And relations \eqref{eq:betagamma} and \eqref{eq:thetathetabar} become
\iea{
  \gamma'=\frac{d_{\rm ls}-d_{\rm ol}}{d_{\rm ls}}\beta',~\bar{\theta}=\frac{d_{\rm ol}}{d_{\rm ls}}\theta. \label{eq:gammapbetapsmall}
}
Substituting into equation \eqref{eq:Ohanianp}, we get the retro-lensing equation
\iea{
  \beta'&=& \frac{d_{\rm ls}+d_{\rm ol}}{d_{\rm ls}-d_{\rm ol}}\theta-\frac{d_{\rm ls}}{d_{\rm ls}-d_{\rm ol}}\Delta\varphi_{mod}'. \label{eq:lensingretrosmall}
}

\subsection{Weak regular lensing}

For relativistic particles in the weak field limit, the deflection angle is given by Eq. \eqref{eq:deltaphiv1finfexp} and the lensing is regular. This deflection angle receives corrections due to velocity at order $\mathcal{O}(1/b)$ and  due to charge at order $\mathcal{O}(1/b^2)$. If we substitute this angle into Eq. \eqref{eq:lensingnormalsmall} and use relation $b=d_{\rm ol}\theta$, the full lensing equation will be a cubic polynomial of $\theta$. The solution of this equation is quite long and it is hard to recognize the effect of charge on the apparent angle.  Knowing that the $\mathcal{O}(1/b^2)$ order correction to $\theta$ is small compared to the $\mathcal{O}(1/b)$ term, rather, we can do an iteration by first solving the truncated quadratic lensing equation and substitute the result into the full lensing equation to solve for corrections. Carrying out this procedure, we obtain two images
\iea{
  \theta_{\pm}&=&\frac{1}{2}\left\{\beta\pm\frac{\sqrt{y}}{v}+ \frac{\pi  \left[3 M^2 \left(v^2+4\right)-Q^2 \left(v^2+2\right)\right] \left(\pm \sqrt{y}-\beta  v \right)}{8M d_{\rm ol} \left(v^2+1\right) \sqrt{y}}\right\},\label{eq:thetaweakv1exppert}
}
where
\ieas{
  y=\beta ^2 v^2 +\frac{8 M  d_{\rm ls}\left(v^2+1\right)}{d_{\rm ol}(d_{\rm ls}+d_{\rm ol})}.
}
The last term in the curl bracket originates from the $\mathcal{O}(1/b^2)$ terms in the deflection angle \eqref{eq:deltaphiv1finfexp}. Therefore, only in this order the effects of charge $Q$ to $\theta$ is present. It is noted that because $\sqrt{y}>\beta v$, this term is positive for $\theta_+$ and negative for $\theta_-$ and therefore make the two images more widely separated. The effect of charge however, is to reduce the amount of this opening because it makes the coefficient for opening smaller. This will be more clearly seen from the lightlike ray case, Eq. \eqref{eq:thetaweakv1exppertv1} or Eq. \eqref{eq:thetaweakvposmneg} and the plots in Fig. \ref{fig:thetaweak}.
Setting $\beta=0$ in Eq. \eqref{eq:thetaweakv1exppert} we get $\theta_{+}=-\theta_{-}$, as expected because of the exact alignment of source, lens and the observer.

\subsubsection{Lightlike ray lensing}

Setting $v=1$ in Eq. \eqref{eq:thetaweakv1exppert} yields the  apparent angle of lightlike ray
\iea{
  \left.\theta_{\pm}\right|_{v=1}&=&\frac{1}{2}\left[\beta\pm\sqrt{y_l}+\frac{3\pi(5M^2-Q^2)(\pm \sqrt{y_l}-\beta)}{16Md_{\rm ol}\sqrt{y_l}}\right], \label{eq:thetaweakv1exppertv1}
}
where
\ieas{
  y_l=y|_{v=1}=\beta^2+\frac{16Md_{\rm ls}}{d_{\rm ol}(d_{\rm ls}+d_{\rm ol})}.
}
Comparing the first two terms and the last term, it is seen that typically, the charge correction to the apparent angle is an order $\mathcal{O}(1/d_{\rm ol})$ smaller than the leading term.

The two images given by Eq. \eqref{eq:thetaweakv1exppertv1} are separated by the angle
\iea{
  \theta_{+}-\theta_{-}=\sqrt{y_l}+\frac{3\pi(5M^2-Q^2)}{16Md_{\rm ol}}. \label{eq:thetaweakvposmneg}
}
The first term is the typical Schwarzschild apparent angular separation when only $\mathcal{O}(1/b)$ order result is taken account into the deflection angle $\Delta\varphi_{mod}$. The second term is order $\mathcal{O}(1/\sqrt{d_{\rm ol}})$ smaller than the first term and includes the effect of charge.

For very small  $\beta$ such that $\beta^2\ll 16Md_{\rm ls}/[d_{\rm ol}(d_{\rm ls}+d_{\rm ol})]$, Eq. \eqref{eq:thetaweakv1exppertv1} can be approximated as
\iea{
  \left.\theta_{\pm}\right|_{v=1}&\approx&\pm \theta_E+c_1\beta+\mathcal{O}(\beta^2), \label{eq:thetaweakv1exppertv1smallbeta}
}
where $\theta_E$ is the angular position of the Einstein ring
\iea{
  \theta_E= \sqrt{\frac{4M  d_{\rm ls} }{d_{\rm ol} (d_{\rm ls}+d_{\rm ol})}}+\frac{3\pi(5 M^2-Q^2)}{32 Md_{\rm ol}},\label{eq:thetaweakE}
}
and
\iea{
   c_1&=& \frac{1}{2}-\frac{3\pi(5M^2-Q^2)}{128M}\sqrt{\frac{(d_{\rm ls}+d_{\rm ol})}{Md_{\rm ls}d_{\rm ol}}}.\label{eq:defc1}
}
We recognize that the first term in Eq. \eqref{eq:thetaweakE} is just the ordinary position of Einstein rings in Schwarzschild spacetime \citep[Eq. (24)]{Jia:2015zon}, and the second term is due to the order $\mathcal{O}(1/b^2)$ correction to the deflection angle.

The magnification of images using particle number conservation is obtained as \cite{Eiroa:2003jf}
\iea{
  \mu_{\pm}&=&\left|\frac{\sin\theta_{\pm}}{\sin\beta}\frac{\dd\theta_{\pm}}{\dd\beta}\right|=\left|\frac{\theta_{\pm}}{\beta}\frac{\dd\theta_{\pm}}{\dd\beta}\right| \\
  &=&\frac{1}{4}\left[1\pm\frac{\sqrt{y_l}}{\beta}+\frac{3\pi(5M^2-Q^2)(\sqrt{y_l}/\beta-1)}{16Md_{\rm ol}\sqrt{y_l}}\right] 
  \left[1\pm\frac{\beta}{\sqrt{y_l}}-\frac{3\pi(5M^2-Q^2)}{16Md_{\rm ol}\sqrt{y_l}}\left(1-\frac{\beta^2}{y_l}\right)\right] \nonumber \\
  &=& \frac{1}{4}\left\{2\pm\left(\frac{\sqrt{y_l}}{\beta}+\frac{\sqrt{\beta}}{\sqrt{y_l}}\right)-\frac{3\pi(5M^2-Q^2)}{16Md_{\rm ol}}\left(\frac{\beta^2}{y_l^{3/2}}-\frac{2}{\sqrt{y_l}}\pm\frac{1}{\sqrt{y_l}}+\frac{1}{\beta}\mp\frac{1}{\beta}\right)\right. \nonumber \\
  &&\left.-\left[\frac{3\pi(5M^2-Q^2)}{16Md_{\rm ol}}\right]^2\left(\frac{1}{\sqrt{y_l}\beta}-\frac{1}{y_l}-\frac{\beta}{y_l^{3/2}}+\frac{\beta^2}{y_l^3}\right)\right\}. \label{eq:muweak}
}
When $\beta$ is very small,  approximating $y_l$ as $y_l\approx 4M d_{\rm ls}/[d_{\rm ol}(d_{\rm ls}+d_{\rm ol})]$, to the $\mathcal{O}(\beta^0)$ order, the magnification becomes
\iea{
  \mu_{\pm}&=&\frac{c_1\theta_{E}}{\beta}\pm c_1^2. \label{eq:muweaksmallbeta}
}
Clearly $\mu_{\pm}$ approaches infinity as for $\beta\to0$. For $\beta>0$ we have $\mu_+>\mu_-$. Moreover, from Eqs.  \eqref{eq:thetaweakE} and \eqref{eq:defc1} we see that as $Q$ increases, $\theta_E$ decreases in the order of $\mathcal{O}(1/d_{\rm ol})$, while $c_1$ increases in the order of $\mathcal{O}(1/\sqrt{d_{\rm ol}})$. After multiplication, for large distances $d_{\rm ls}$ and $d_{\rm ol}$, it can be shown that the magnification \eqref{eq:muweaksmallbeta} decreases as $Q$ increases.

\subsubsection{Velocity correction}

For relativistic timelike ray, we can expand Eq. \eqref{eq:thetaweakv1exppert} to yield the velocity correction to the apparent angles to the first order of $(1-v)$
\iea{
  &&\theta_{\pm}= \left.\theta_{\pm}\right|_{v=1}+(1-v)\left(\pm d_1+ d_{1}'\beta\right),\label{eq:thetaweakv1exppertv1exp}
}
where $\theta_{\pm}$ is given by Eq. \eqref{eq:thetaweakv1exppertv1smallbeta} and
\iea{
  d_{1}&=&\frac{4M d_{\rm ls}}{  d_{\rm ol}(d_{\rm ls}+d_{\rm ol})\sqrt{y_l}} + \frac{\pi \left(9 M^2-Q^2\right)}{32 M d_{\rm ol}},\label{eq:defd1}\\
  d_1'&= &-\frac{\pi \left(9 M^2-Q^2\right)}{32 M d_{\rm ol}\sqrt{y_l}}+\frac{3 \pi  d_{\rm ls} \left(5 M^2-Q^2\right)}{4 (d_{\rm ls}+d_{\rm ol})d_{\rm ol}^2 y_l^{3/2} }.\label{eq:defd1p}
}
Noting that $\sqrt{y_l}$ is a small quantity, the dominate velocity correction to the apparent angle comes from the first term in $d_1$. Therefore, as $v$ deviate from $1$ the magnitude of both $\theta_+$ and $\theta_-$ increases. Moreover, comparing to the charge correction to the apparent angle given by the last term in \eqref{eq:thetaweakv1exppert}, the velocity correction is order $\mathcal{O}(\sqrt{d_{\rm ol}})$ higher, just as one would anticipate from their contribution to the deflection angle in Eq. \eqref{eq:deltaphiv1finfexpv1vcorre}.

The magnification for timelike rays with very small $\beta$ also receives a correction due to velocity as
\iea{
  \mu_{\pm}(v)&=& \frac{c_1\theta_{E}}{\beta}\pm c_1^2+\left(\frac{c_1d_{1}\pm\theta_Ed_1'}{\beta}\pm2c_1d_1'\right)(1-v).
}
Since $d_1$ is about the same order as $d_1'$ and $c_1\gg\theta_E$ which $\theta_E$ is of order $\mathcal{O}(1/\sqrt{d_{\rm ol}})$, for small enough $\beta$ we see that the coefficient of $(1-v)$ is always positive. Consequently,  the deviation of $v$ from $1$ increase both $\mu_+$ and $\mu_-$.

\begin{figure}[htp]
  \centering
  \subfloat[]{\label{fig:thetaweakv1}\includegraphics[width=0.45\textwidth]{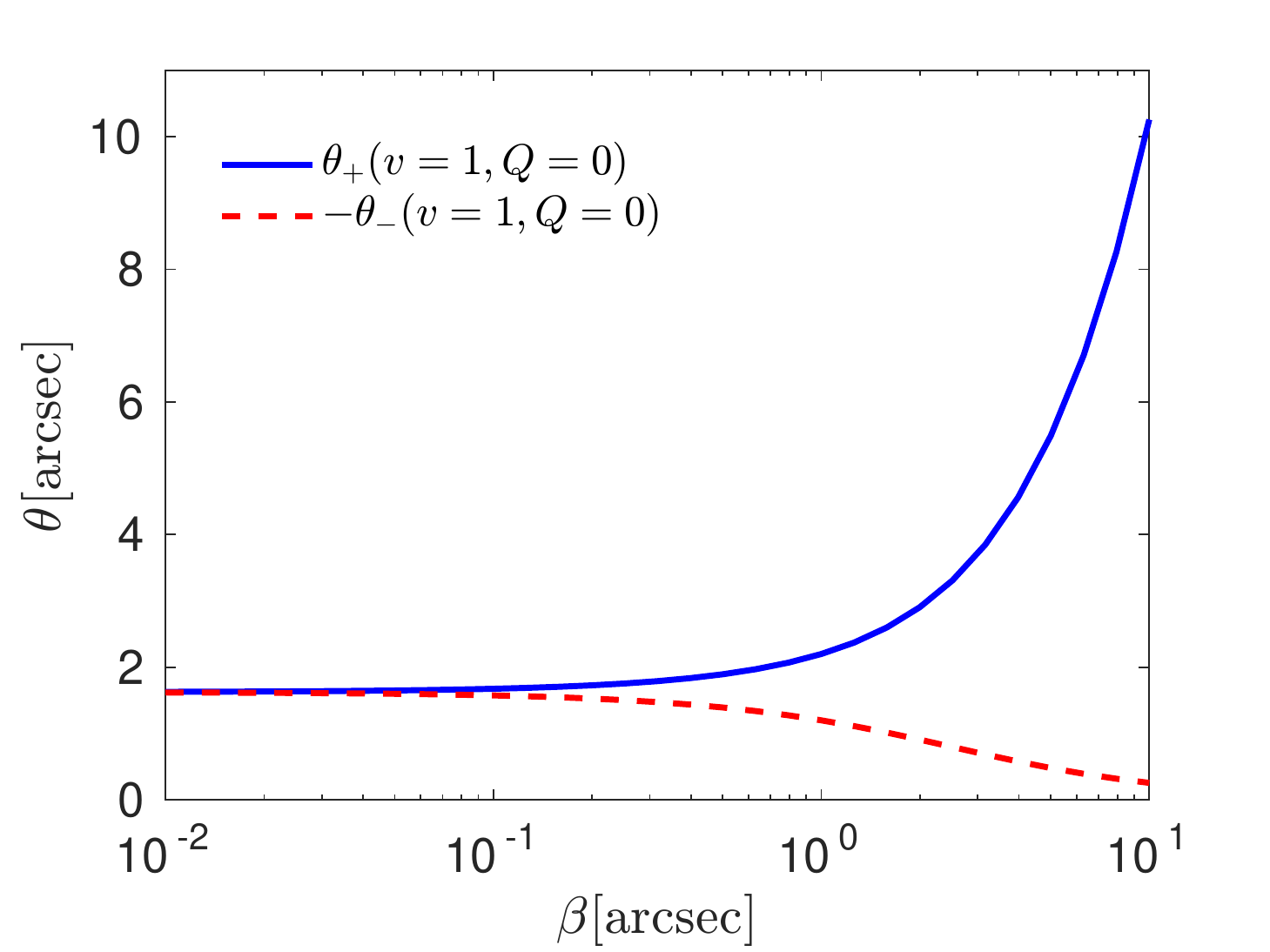}}\\
  \subfloat[]{\label{fig:thetaweakbeta10q0}\includegraphics[width=0.45\textwidth]{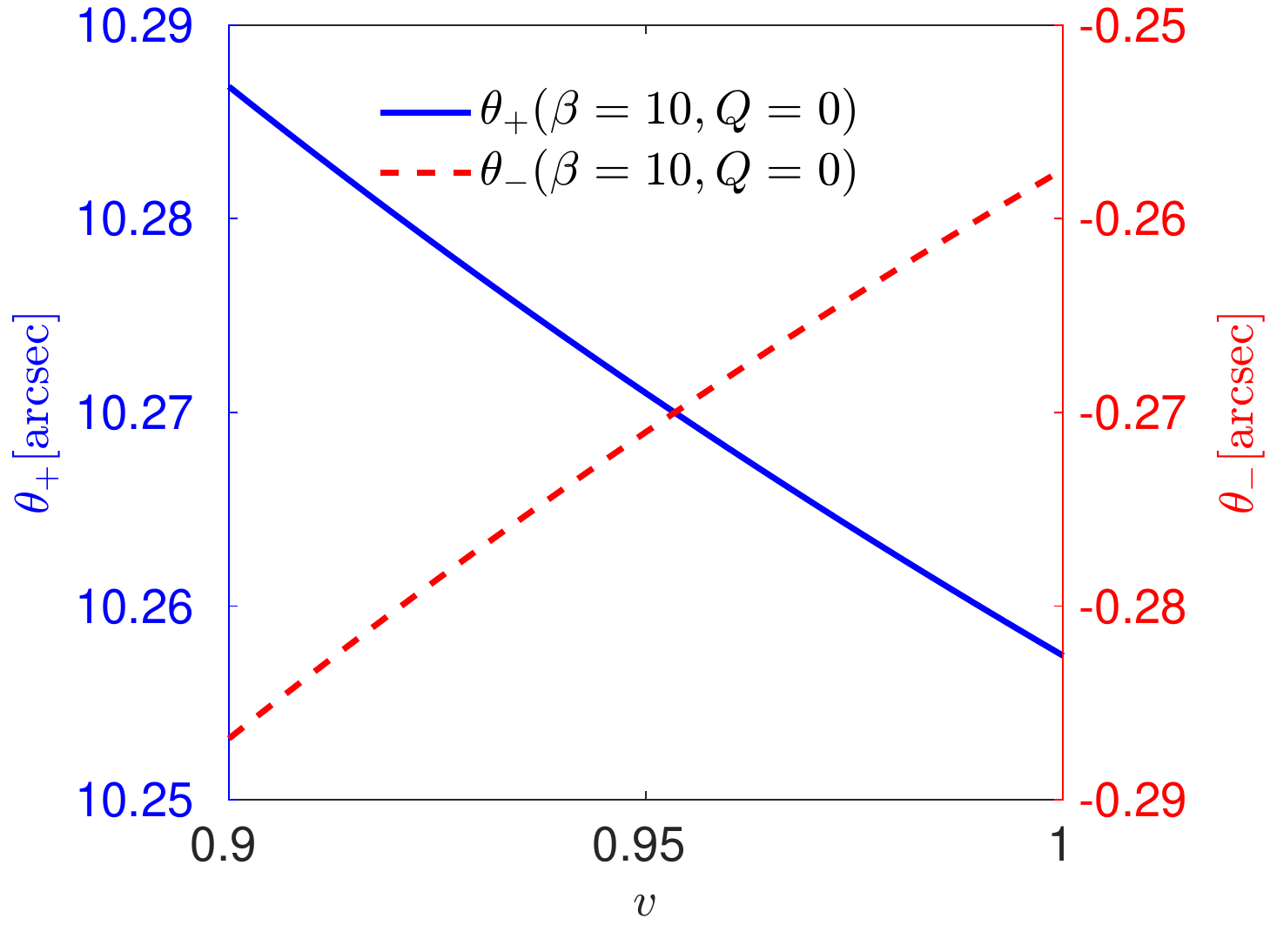}}~
  \subfloat[]{\label{fig:thetaweakbeta10v1}\includegraphics[width=0.45\textwidth]{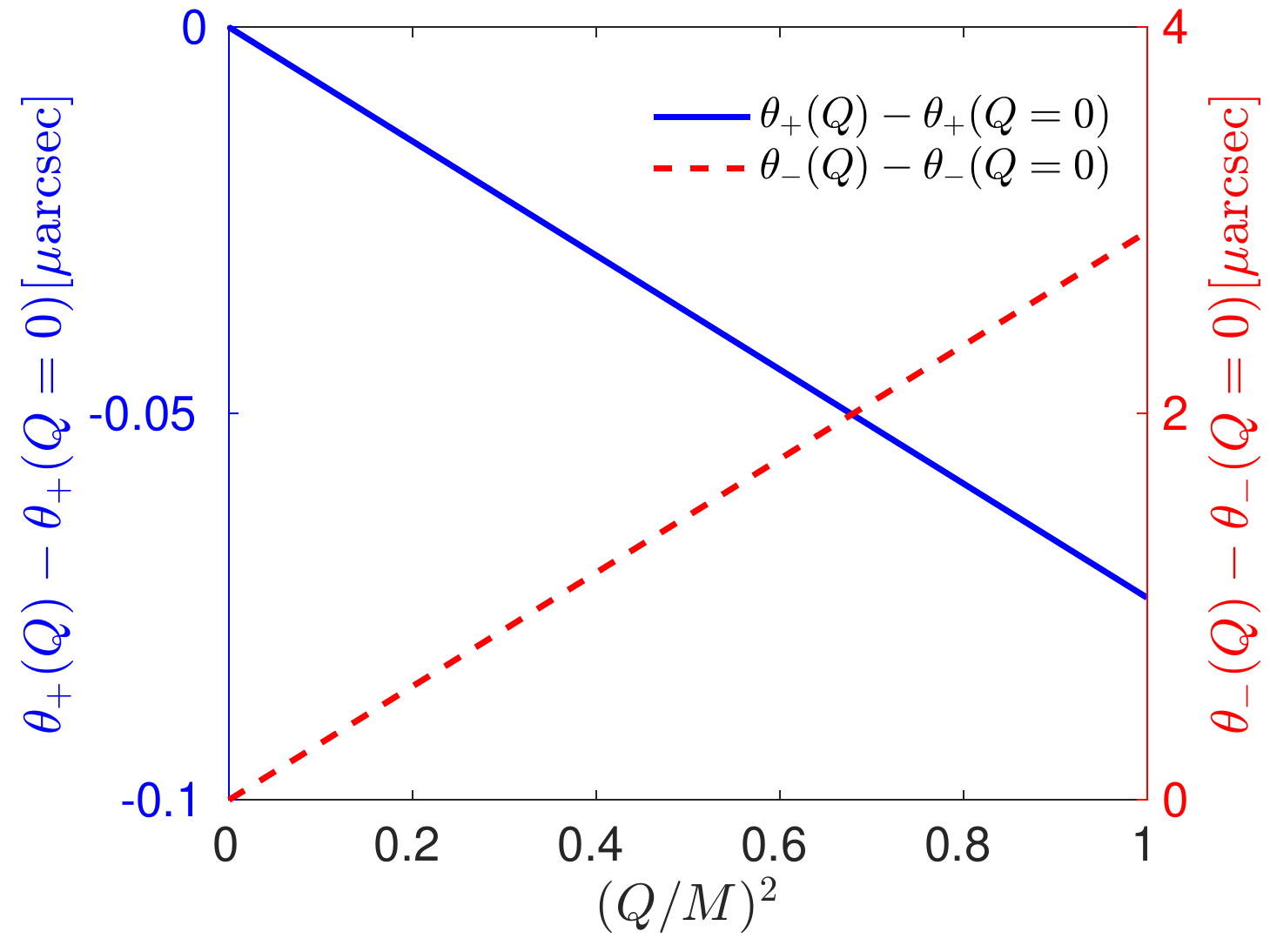}}\\
  \caption{The apparent angle of $\theta$ as a function of $\beta$, using Eq. \eqref{eq:thetaweakv1exppert}. Fig. \ref{fig:thetaweakv1}: $\theta_\pm (Q=0,~v=1,~\beta)$; Fig. \ref{fig:thetaweakbeta10q0}: $\theta_\pm (Q=0,~v,~\beta=10~{\rm arcsec})$;  Fig. \ref{fig:thetaweakbeta10v1}: $\theta_\pm (Q,~v=1,~\beta=10~{\rm arcsec})$. Other parameters are $M=4.31\times 10^6 M_{\odot}~\text{\cite{Gillessen:2008qv}},~d_{\rm ls}=13.9~{\rm kpc},~d_{\rm ol}=8.33~{\rm kpc}$.}
  \label{fig:thetaweak}
\end{figure}

In Fig. \ref{fig:thetaweak} we plot the apparent angle as a function of $\beta$, $Q$ or $v$ when other parameters are fixed. We assumed that the lensing object is the Sgr A* in the center of the Galaxy and the source is located on the edge of the Galaxy stellar disk and on the opposite side of our solar system. In general, the reduction effects of charge and the increasing effect of velocity deviation on the magnitude of $\theta_+$ and $\theta_-$ are confirmed in Fig. \ref{fig:thetaweakbeta10q0} and Fig. \ref{fig:thetaweakbeta10v1}. Moreover, we also note by comparing Fig. \ref{fig:thetaweakbeta10q0} and Fig. \ref{fig:thetaweakbeta10v1} that the effect of charge is much smaller than that of velocity as we have found previously by inspecting Eq. \eqref{eq:thetaweakv1exppertv1exp}.

\subsection{Strong regular lensing} \label{sec:strongregular}

In the strong field limit, the deflection angle \eqref{eq:deltaphiv1fcexp} indicates that a particle can loop around the center many times and travel out in a forward direction to form a regular lensing geometry. We will consider this scenario in this subsection.

\subsubsection{Lightlike ray lensing}

The deflection angle expansion \eqref{eq:deltaphiv1fcexp} for massive particles is too complicated to allow the lensing equation solved for general $v$. Thus, we start by setting $v=1$ and solve the apparent angle of lightlike rays first, and then find the velocity correction to it.

For the regular lensing in the strong field limit,
\iea{
  \Delta\varphi_{mod}=\Delta\varphi-(2n+1)\pi,~(n=1,~2,~3,~\cdots).
}
Substituting into lensing Eq. \eqref{eq:lensingnormalsmall} and using the deflection angle expansion \eqref{eq:deltaphiv1fcexpv1b}, replacing $\delta_b$ by $b-b_c$ and then $b$ by $d_{\rm ol}\theta$, we can work out the apparent angle of the images on two sides of the lens axis
\iea{
  \theta_{n,\pm}&=& \pm\left\{\sqrt{\frac{x^2+2}{2(x^2-1)}}\frac{d_{\rm ls}W[g_n(\pm\beta,h)]}{ (d_{\rm ls}+d_{\rm ol})}+\frac{M \left(x^2+2\right)^{3/2}}{\sqrt{2} x d_{\rm ol}} \right\},~(n,=1,2,3,\cdots),\label{eq:thetav1strongexa}
}
with $W$ being the Lambert-$W$ function, $x$ given by Eq. \eqref{eq:xdef}, and
\iea{
  g_n(\beta,h)&=& \frac{128\left(x^2-1\right)^{7/2} \left(x^2+2\right)}{x^5 \left(\sqrt{x^2-1}+x\right)^2}\frac{ (d_{\rm ls}+d_{\rm ol})M  }{ d_{\rm ls} d_{\rm ol}}\nonumber \\
  &&\times\exp\left(-\sqrt{\frac{x^2-1}{x^2+2}} \left\{\sqrt{2} \pi  (2 n+1)+\frac{(d_{\rm ls}+d_{\rm ol}) \left[M \left(x^2+2\right)^{3/2}-\sqrt{2} \beta  x d_{\rm ol}\right]}{x d_{\rm ls} d_{\rm ol}}\right\}\right). \label{eq:gnbeta}
}
When $h=0$ we have $x=2$ and Eq. \eqref{eq:thetav1strongexa} simplifies to
\iea{
  \theta_{n,\pm}=\pm\left\{\frac{d_{\rm ls}W[g_n(\pm\beta,h=0)]}{d_{\rm ls}+d_{\rm ol}}+\frac{3\sqrt{3}M}{d_{\rm ol}}\right\},~(n=1,~2,~3,~\cdots).
}
This equation is  identical to equation (33) in our previous paper \cite{Jia:2015zon}.

The magnification corresponding to Eq. \eqref{eq:thetav1strongexa} is given by
\iea{
  \mu_{n,\pm}&=& \frac{\theta_\pm}{\beta}\frac{W[g_n(\pm\beta,h)]}{1+W(g_n(\pm\beta,h))}, ~(n=1,~2,~3,~\cdots),
}
which has the same form as equation (36) in \cite{Jia:2015zon} with $g_n(\beta)$ replaced by $g_n(\beta,h)$ due to the presence of charge.

Even though Eq. \eqref{eq:thetav1strongexa} is exact, the Lambert$-W$ function in it makes velocity corrections difficult to study. Therefore, we attempt to approximate the apparent angle as was done in Refs. \cite{Eiroa:2003jf,Jia:2015zon}. The logic here is that the apparent angle $\theta_n$ corresponding to a small $\Delta\varphi_{mod}$ should be close to an apparent angle $\theta_{n,0}$ correspond to $\Delta\varphi_{mod}=0$, and therefore $\theta_n$ can be expanded around this value. The value of $\theta$ at $\Delta\varphi_{mod}=0$, i.e., $\Delta\varphi=(2n+1)\pi$, is easily find by using Eq. \eqref{eq:deltaphiv1fcexpv1} and again $b=d_{\rm ol}\theta$ to be
  \iea{
	\theta_{n,0}&=&\frac{M \left(x^2+2\right)^{3/2}}{\sqrt{2} x d_{\rm ol}} \left(1+\xi_n\right),~(n=1,~2,~3,~\cdots), \label{eq:thetastrongv1lead}
  }
  where
  \iea{
	\xi_n&=& \frac{128 \left(x^2-1\right)^3 e^{-\sqrt{2}(2 n+1)  \pi \sqrt{\frac{x^2-1}{x^2+2}}}}{x^4 \left(\sqrt{x^2-1}+x\right)^2}. \label{eq:defxin}
  }
The next step, which is also the key step, is to relate the actual (not necessarily zero) $\Delta\varphi_{mod}$ with $(\theta-\theta_{n,0})$ so that $\Delta\varphi_{mod}$ can be used in the lensing equation \eqref{eq:lensingnormalsmall} and then $\theta$ can be solved. For this purpose, replacing $\delta_b$ in Eq. \eqref{eq:deltaphiv1fcexpv1b} by $b-b_c$ and then $b$ by $ d_{\rm ol}\theta$ and finally $\theta$ by $\theta_{n,0}+(\theta-\theta_{n,0})$. One can expand $\Delta\varphi$ in \eqref{eq:deltaphiv1fcexpv1b} to the first order of $(\theta-\theta_{n,0})$. This yields
  \iea{
	\Delta\varphi&=& (2n+1)\pi+\frac{ x d_{\rm ol} \left(\theta -\theta _{n,0}\right)}{M\sqrt{x^2-1} \left(x^2+2\right)\xi_n}, ~(n=1,~2,~3,~\cdots).
  }
Furthermore, because $\Delta\varphi=(2n+1)\pi+\Delta\varphi_{mod}$, we obtain the desired relation between $\Delta\varphi_{mod}$ and $\theta-\theta_{n,0}$
  \iea{
	\Delta\varphi_{mod}&=& \frac{ x d_{\rm ol} \left(\theta -\theta _{n,0}\right)}{M\sqrt{x^2-1} \left(x^2+2\right)\xi_n}.
  }
Finally, substituting this $\Delta\varphi_{mod}$ into the lensing equation \eqref{eq:lensingnormalsmall}, we can solve the apparent angle $\theta$ as
\iea{
  \theta_{n,\pm}(\beta)&=& \pm\left[\theta _{n,0}+ \frac{M\sqrt{x^2-1}\left(x^2+2\right)(d_{\rm ls}+d_{\rm ol}) \xi_n\left(\pm\beta -\theta _{n,0}\right) }{ x d_{\rm ol} d_{\rm ls}}\right],~(n=1,~2,~3,~\cdots). \label{eq:thetastrongv1}
}

In Eq. \eqref{eq:thetastrongv1}, since $x=\sqrt{\sqrt{9-8h^2}+1}$ will decrease as $Q$ increases, all the numerators and denominators in $\theta_{n,0},~\xi_n$ and last term in the bracket of \eqref{eq:thetastrongv1} decrease. Therefore, in order to recognize the effect of charge to the apparent  angle in the strong regular lensing case, we should further expand the apparent angle in the small $Q$ limit. However, an even simpler way to recognize this is from the effect of charge to deflection angle in \eqref{eq:deltaphiv1fcexpvcorrech0exp}. Since the increase of $Q$ will decrease the deflection angle, one would expect that the apparent angle will decrease too in the strong lensing case, just as in the weak lensing one. This is indeed confirmed in the plot Fig. \ref{fig:thetastrongposbeta10n1}.

In the limit of large $n$, we have $\xi_n\to0$ and therefore the apparent angle \eqref{eq:thetastrongv1} reduces to
\iea{
  |\theta_{\infty,\pm}|=\frac{M(x^2+2)^{3/2}}{\sqrt{2}xd_{\rm ol}}. \label{eq:thetastrongv1Eninf}
}
Under this limit the lightlike ray will loop infinitely many circles around the gravitational center. Therefore, one would expect that this limiting ray is the one that approaches and leaves the gravitational center with impact parameter equaling to the critical one $b_c(h,v=1)$. Geometrically this implies
\iea{
|\theta_{\infty,\pm}|=\frac{b_c(h,v=1)}{d_{\rm ol}}. \label{eq:thetaneq}
}
Actually, this equation can be nicely shown to equal result \eqref{eq:thetastrongv1Eninf} if in the later the definition of $x$ and $b_c$ (i.e., Eq. \eqref{eq:bchv1}) are used.  This limiting value is also confirmed by Fig. \ref{fig:thetastrongposnegv1q0beta10}.

Corresponding to apparent angle \eqref{eq:thetastrongv1} we have the magnification
\iea{
  \mu_{n,\pm}&=& \frac{(1-c_2)c_2\theta_{n,0}}{\beta}\pm c_2^2,~(n=1,~2,~3,~\cdots). \label{eq:muregular}
}
  where
\iea{
	c_2&=& \frac{M\sqrt{x^2-1}(x^2+2)(d_{\rm ls}+d_{\rm ol})\xi_n}{xd_{\rm ls}d_{\rm ol}}. \label{eq:defc2}
}
This magnification of the relativistic images can be compared to the weak lensing magnification, Eq. \eqref{eq:muweaksmallbeta}. First noting that $\xi_n\ll1$ for $n\geq1$, $\theta_{n,0}$ in Eq. \eqref{eq:thetastrongv1lead} can be approximated by $\theta_{\infty,\pm}$ in Eq. \eqref{eq:thetastrongv1Eninf}. In addition, for large distances $d_{\rm ls}$ and $d_{\rm ol}$, we have $(d_{\rm ls}+d_{\rm ol})/(d_{\rm ol}d_{\rm ls})\ll1$ and consequently from Eqs. \eqref{eq:defc1} and \eqref{eq:defc2} we see that $c_1\approx1/2$ and  $c_2\ll 1$.  When $\beta$ is very small, the second terms in \eqref{eq:muweaksmallbeta} and \eqref{eq:muregular} can be dropped, and finally the ratio between magnifications of regular lensed images in the strong field limit and the weak field limit, and that between different orders of strongly lensed image become respectively
\iea{
  \frac{\mu_{n,\pm}}{\mu_{\pm}}&\approx& \frac{c_2\theta_{n,0}/\beta}{\theta_E/(2\beta)}
  \approx\frac{\sqrt{M}(x^2+2)^{5/2}\sqrt{x^2-1}}{\sqrt{2}x^2}\left(\frac{d_{\rm ls}+d_{\rm ol}}{d_{\rm ls}d_{\rm ol}}\right)^{3/2}\xi_n
  \ll1, ~(n=1,~2,~3,~\cdots),\label{eq:musl1}\\
  \frac{\mu_{n+1}}{\mu_n}&\approx& \frac{c_2(n+1)\theta_{n+1,0}/\beta}{c_2(n)\theta_{n,0}/\beta}
  \approx e^{-2\sqrt{2}\pi\sqrt{\frac{x^2-1}{x^2+2}}}<1,~(n=1,~2,~3,~\cdots).\label{eq:musl2}
}
Eq. \eqref{eq:musl1} shows that the magnification of relativistic images is much smaller than that of the weakly lensed images. And Eq. \eqref{eq:musl2} suggests that the magnification of relativistic images roughly form a geometric series with a constant ratio that is determined by $Q$ and smaller than 1. In addition,
for Schwarzschild case $Q=0$, we have $x=2$, and
\iea{
  \frac{\mu_{n+1,\pm}}{\mu_{n,\pm}}\approx e^{-2\pi} ,
}
while for extremal case $Q=M$ and consequently $x=\sqrt{2}$, then
\iea{
  \frac{\mu_{n+1,\pm}}{\mu_{n,\pm}}\approx e^{-\sqrt{2}\pi}.
}
We see that the increase of charge decreases makes the geometric series decrease slower by making it common factor larger.

\subsubsection{Velocity corrections}

\begin{figure}[htp]
  \centering
  \subfloat[]{\label{fig:thetastrongposnegv1q0n1}\includegraphics[width=0.45\textwidth]{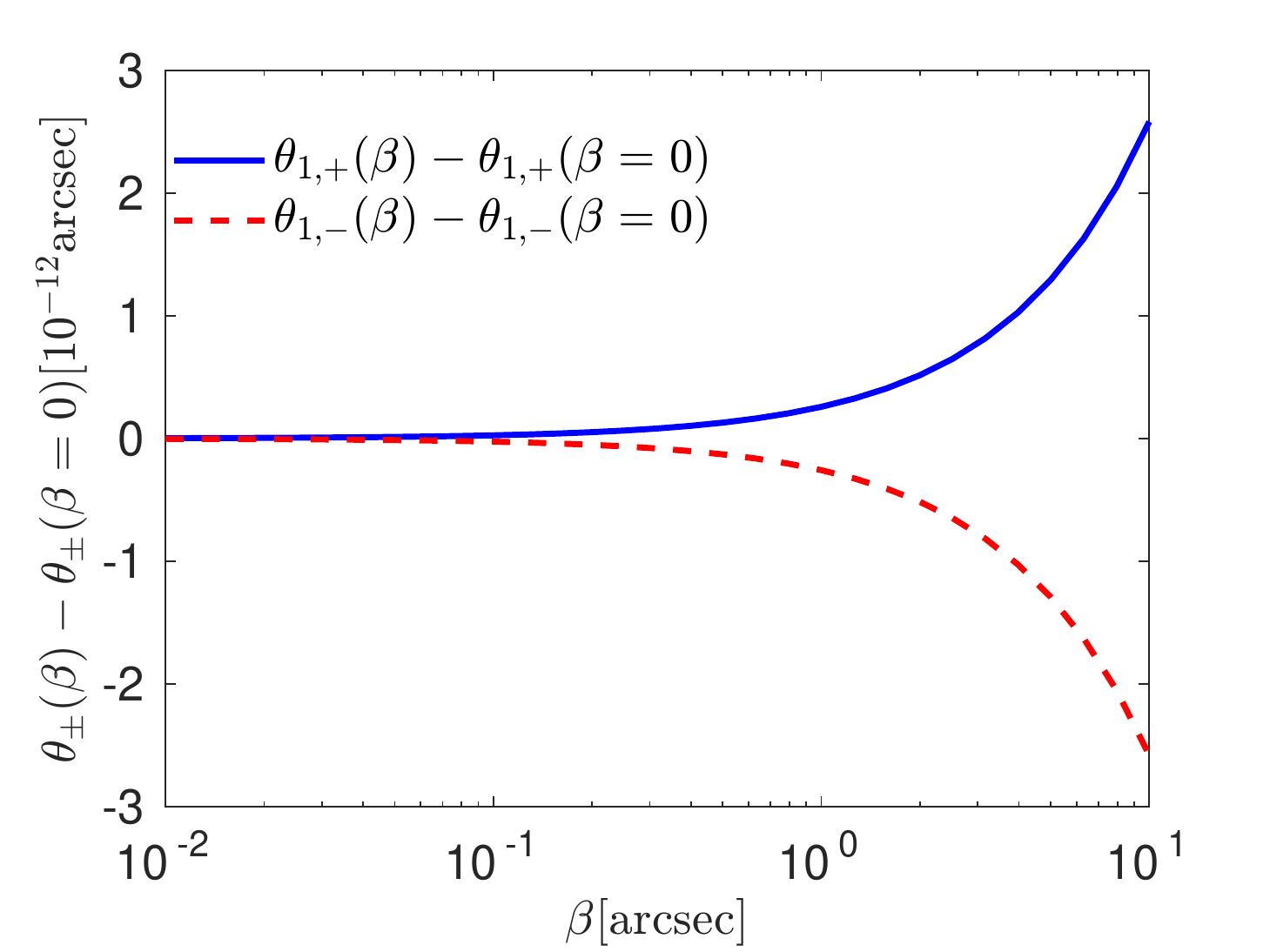}}\\
  \subfloat[]{\label{fig:thetastrongposnegv1q0beta10}\includegraphics[width=0.45\textwidth]{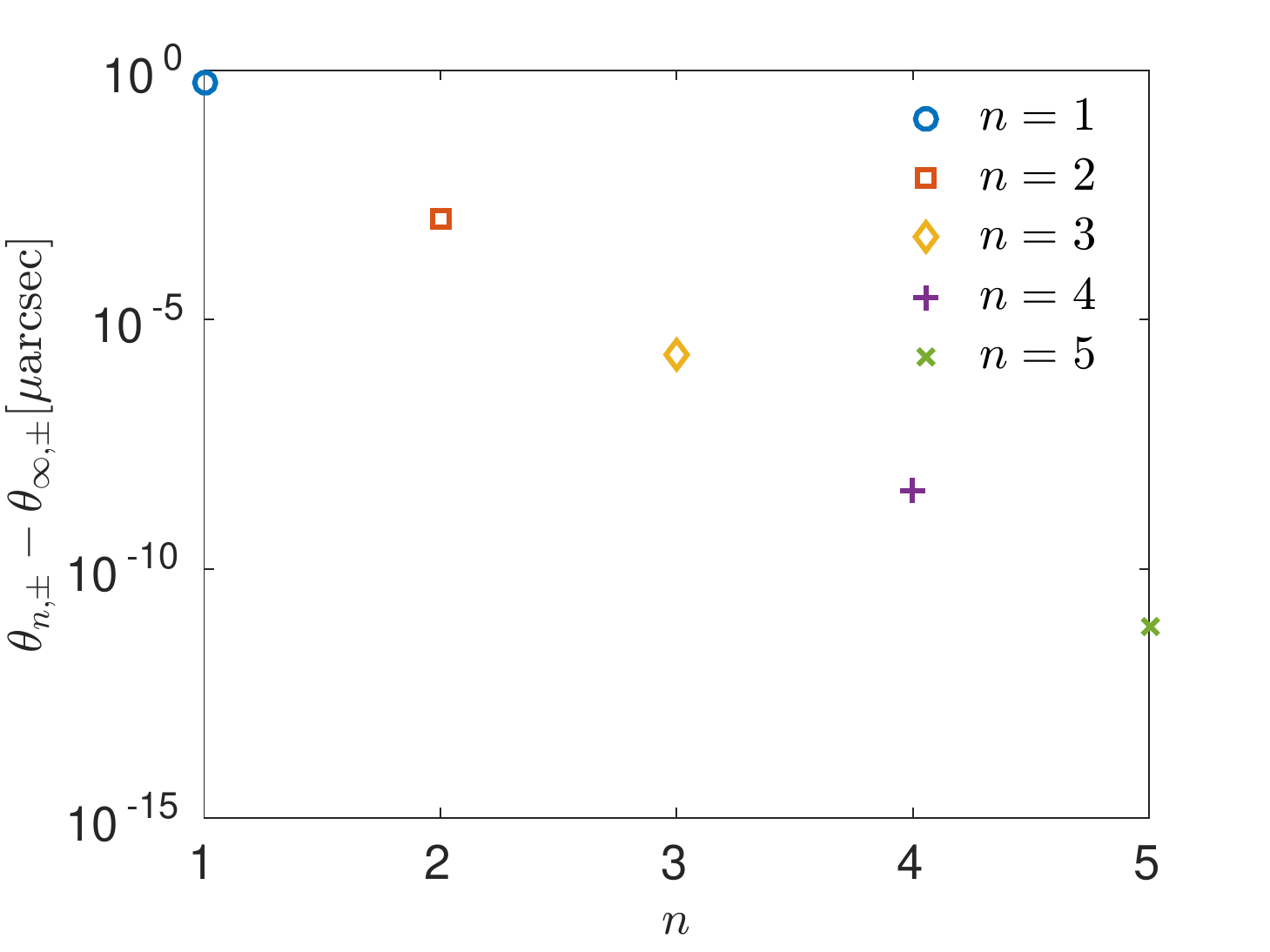}}~
  \subfloat[]{\label{fig:thetastrongposbeta10n1}\includegraphics[width=0.45\textwidth]{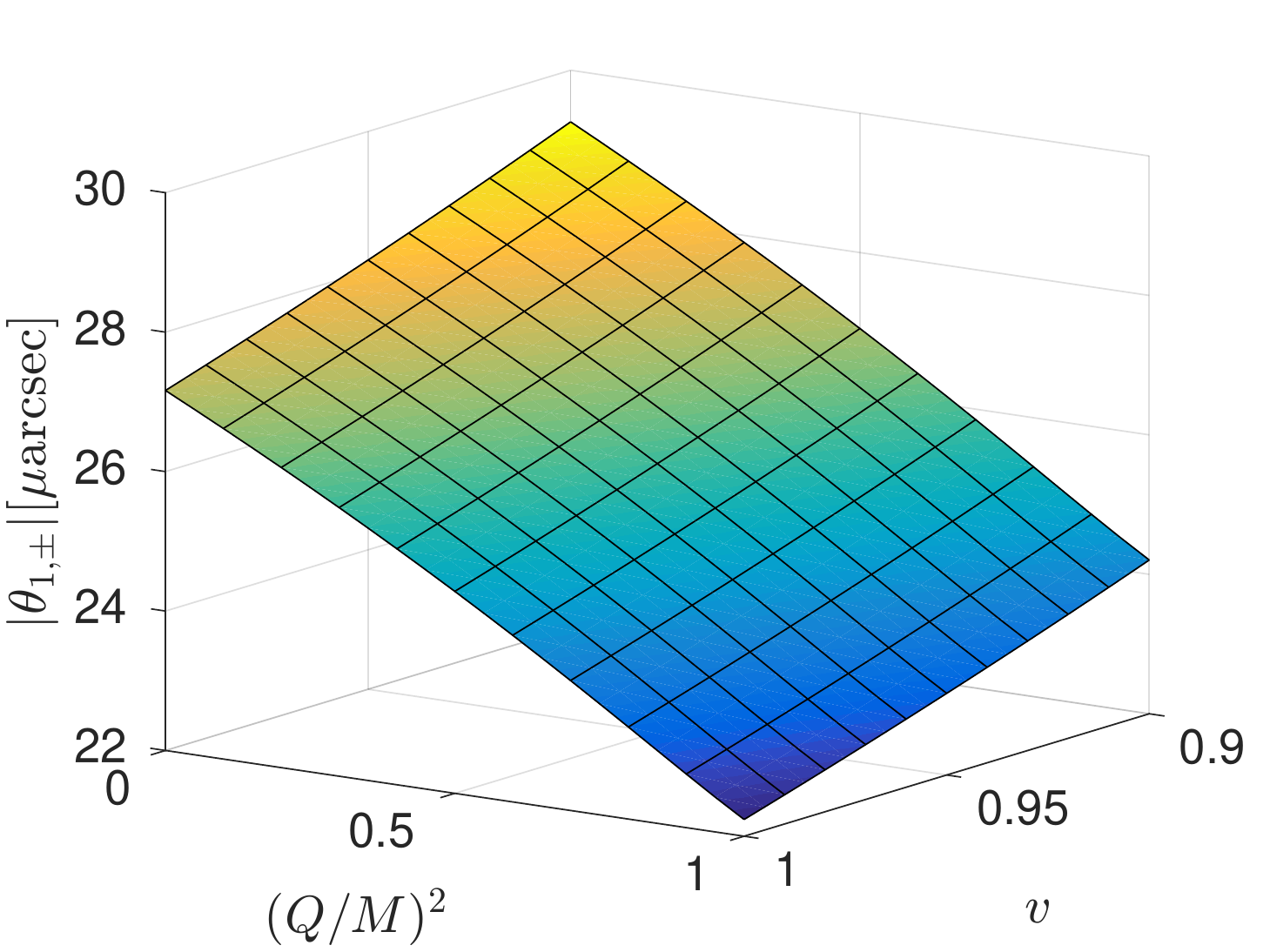}}\\
  \caption{Apparent angle of relativistic images.  Fig. \ref{fig:thetastrongposnegv1q0n1}: $\theta_{1,\pm}(\beta,~h=0,~v=1)$, where $|\theta_{1,\pm}(\beta=0)|=26.6308~\mathrm{\mu arcsec}$; Fig. \ref{fig:thetastrongposnegv1q0beta10}: $\theta_{n,\pm}(\beta=10~{\rm arcsec},h=0,v=1)$; Fig. \ref{fig:thetastrongposbeta10n1}: $\theta_{1,\pm}(\beta=10~{\rm arcsec},h,v)$.  The $\theta_{n,\pm}$ in Fig. \ref{fig:thetastrongposnegv1q0beta10} and \ref{fig:thetastrongposbeta10n1} are numerically so close that they do not distinguish. Other parameters are the same as in Fig. \ref{fig:thetaweak}.}
  \label{fig:thetastrong}
\end{figure}

Following the expansion procedure used in the Eqs. \eqref{eq:thetastrongv1lead}$-$\eqref{eq:thetastrongv1}, the apparent angle for relativistic particles with velocity $v$ close to $1$ now becomes
\iea{
	\theta_{n,\pm}(\beta,v)&=&\pm\left\{ \theta _{n,0}(h,v)+\frac{M \sqrt{x^2-1}(x^2+2) (d_{\rm ls}+d_{\rm ol}) \xi_n(h,v)\left[\pm\beta -\theta _{n,0}(h,v)\right]}{x  d_{\rm ol}d_{\rm ls}(1+a_2)v}\right\},~(n=1,~2,~3,~\cdots). \label{eq:thetastrongv1correc}
  }
Here
\iea{
  a_2&=& \frac{(2x^4-x^2-4)x^2(1-v)}{8(x^2-1)^2(x^2+2)}.\label{eq:defa2}
}
$\theta_{n,0}$ and $\xi_n$ now receive a velocity correction
  \iea{
	\theta_{n,0}(h,v)&=&\frac{M \left(x^2+2\right)^{3/2}}{\sqrt{2} x d_{\rm ol}v} \left[1+\xi_n(h,v)\right] -\frac{M x \sqrt{x^2+2} }{2 \sqrt{2}x d_{\rm ol} v }(1-v),~(n=1,~2,~3,~\cdots), \label{eq:thetan0v} \\
	\xi_n(h,v)&=& \frac{128 \left(x^2-1\right)^3 e^{-z}}{x^4 \left(\sqrt{x^2-1}+x\right)^2},~(n=1,~2,~3,~\cdots), \label{eq:xinv}
	}
with
\iea{ z&=&\frac{1}{1+a_2}\left[(2n+1)\sqrt{2}\pi\sqrt{\frac{x^2-1}{x^2+2}} +\frac{a'_1(1-v)}{4(x^2-1)^2(x^2+2)}\right],\label{eq:thetan0vz}
  }
and $a'_1$ is a function of $x$ defined in Eq. \eqref{eq:defa1p}. Further expanding $v$ around $1$, one should easily find the apparent angle $\theta$ to the order of $\mathcal{O}(1-v)$. The expression is quite long and not shown here. One important point here is that the coefficient of $(1-v)$ term can be proven to be positive (negative) for $\theta_+(\theta_-)$. Therefore, similar to the case of weak lensing, deviation of velocity from $1$ will enlarge the magnitude of $\theta_{n,\pm}$. This is confirmed in plot Fig. \ref{fig:thetastrongposbeta10n1}.  Again, letting $n\to\infty$, we see that $\xi_n(h,v)\to0$ too, then Eq. \eqref{eq:thetastrongv1correc} becomes
  \iea{
	|\theta_{\infty,\pm}(\beta,v)|= \frac{M(x^2+2)^{3/2}}{\sqrt{2}xd_{\rm ol}v}-\frac{Mx\sqrt{x^2+2}}{2\sqrt{2}d_{\rm ol}v}(1-v)\approx \frac{M(2+x^2)^{3/2}}{\sqrt{2}xd_{\rm ol}}+\frac{M\sqrt{x^2+2}(x^2+4)}{2\sqrt{2}xd_{\rm ol}}(1-v). \label{eq:thetastrongv1Eninfvcorrec}
  }
Comparing to Eq. \eqref{eq:thetastrongv1Eninf} we see that the deviation of $v$ from $1$ increases the asymptotic value of the apparent angle of relativistic images.

The magnification corresponding to \eqref{eq:thetastrongv1correc} can be found as
\iea{
  \mu_{n,\pm}(v)&=& \frac{(1-d_2)d_2\theta_{n,0}(h,v)}{\beta}\pm d_2^2,~(n=1,~2,~3,~\cdots),
}
where
\iea{
  d_2&=& \frac{M\sqrt{x^2-1}(x^2+2)(d_{\rm ls}+d_{\rm ol})\xi_n(h,v)}{xd_{\rm ol}d_{\rm ls}(1+a_2)v}.
}

In Fig. \ref{fig:thetastrong}, we show the dependency of apparent angle $\theta_{n,\pm}$ on $\beta,~h,~v$ and $n$ in the strong regular lensing case. First of all, one sees that even the first relativistic image ($n=1$) and certainly higher order ($n\geq2$) images are all in the order of micro-arcsec, much smaller than the arcsec order image in the weak regular lensing case in Fig. \ref{fig:thetaweak}. This is similar to the Schwarzschild case studied in \cite{Jia:2015zon} and can be understand from the fact that for strong lensing $\theta_{n,\pm}\lesssim b_n/d_{\rm ol}\ll \theta_{\pm}$, where $b_n$ is the impact parameter at which
\iea{
  \Delta\varphi(h,v,b_n)&=& (2n+1)\pi, ~(n=1,~2,~3,~\cdots) \nonumber
}
and $b_n$ is always bounded above (see Fig. \ref{fig:rcfcbc}). Secondly, from Fig. \ref{fig:thetastrongposnegv1q0beta10} one sees that there exist for any fixed $\beta$ and $v$ an asymptotic value for $\theta_{n,\pm}(\beta,v)$ as $n$ increases, which is consistent with our observation in Eq. \eqref{eq:thetastrongv1Eninf}. The value of these apparent angles for $n$ from 1 to 5 are listed in Table \ref{tab:thetaforn}. Finally, for the effect of $Q$ and $v$, we see from Fig. \ref{fig:thetastrongposbeta10n1} that in general the effects of increasing $Q$ and decreasing $v$ becomes comparable for relativistic images for the parameters ($d_{\rm ol},~d_{\rm ls}$ and $\beta$) we studied. This is in contrast to the weak lensing where the effect of $v$ is orders larger than that of $Q$.

\subsection{Retro-lensing} \label{sec:strongretro}

For the retro-lensing equation given by Eq. \eqref{eq:lensingretrosmall}, we see that it is equivalent to the regular lensing Eq. \eqref{eq:lensingnormalsmall} if we replace in the latter $\beta$ by $(d_{\rm ls}-d_{\rm ol})\beta'/(d_{\rm ls}+d_{\rm ol})$, and $\Delta\varphi_{mod}$ by $\Delta\varphi_{mod}'$. The last replacement is indeed just a replacement of $(2n+1)\pi$ by $2n\pi$ in the calculations in subsection \ref{sec:strongregular}. Therefore, the solution to the apparent angle in Eq. \eqref{eq:lensingretrosmall} can be directly obtain by doing the same replacement in the solutions to Eq. \eqref{eq:lensingnormalsmall}. For lightlike rays, after doing the replacement to solution \eqref{eq:thetastrongv1}, one obtains their apparent angle in the retro-lensing case
\iea{
	\theta'_{n,\pm}(\beta')&=& \pm\left[\theta '_{n,0}+\frac{M\sqrt{x^2-1}(x^2+2)(d_{\rm ls}-d_{\rm ol})\xi'_{n}\left(\pm\beta' -\frac{d_{\rm ls}+d_{\rm ol}}{d_{\rm ls}-d_{\rm ol}} \theta '_{n,0}\right)  }{x  d_{\rm ol} d_{\rm ls}}\right],~(n=1,~2,~3,~\cdots), \label{eq:thetav1retro}
}
where $\xi'_{n}$ is equivalent to $\xi_n$ in Eq. \eqref{eq:defxin} but with $(2n+1)\pi$ replaced by $2n\pi$,
\iea{
	\xi'_n&=& \frac{128 \left(x^2-1\right)^3 e^{-2\sqrt{2} n  \pi \sqrt{\frac{x^2-1}{x^2+2}}}}{x^4 \left(\sqrt{x^2-1}+x\right)^2},~(n=1,~2,~3,~\cdots), \label{eq:defxinp}
}
and $\theta'_{n,0}$ is equivalent to $\theta_{n,0}$ in Eq. \eqref{eq:thetastrongv1lead} with $\xi_n$ replaced by $\xi'_{n}$. Note that for large $n$ we also have $\xi'_n$ approaches $0$ as $\xi_n$ does and consequently the limiting value of $\theta'_{n,\pm}$ is also given by Eq. \eqref{eq:thetastrongv1Eninf}.

The magnification corresponding to Eq. \eqref{eq:thetav1retro} is given by
\iea{
  \mu'_{n,\pm}&=& \frac{\left(1-\frac{d_{\rm ls}+d_{\rm ol}}{d_{\rm ls}-d_{\rm ol}}c'_{2}\right)|c'_{2}|\theta'_{n,0}}{\beta'}\pm c_{2}^{\prime 2}, ~(n=1,~2,~3,~\cdots),\label{eq:muretro}
}
where
\iea{
  c'_{2}&=&\frac{M\sqrt{x^2-1}(x^2+2)(d_{\rm ls}-d_{\rm ol})\xi'_{n} }{x  d_{\rm ol} d_{\rm ls}}.
}
By the assumptions that distances $d_{\rm ls}$ and $d_{\rm ol}$ are large and noting $\xi'_n\ll1$, we see that $|c'_{2}|\ll1$ too. Therefore, in the limit $\beta'\to0$ the $c_2^{\prime 2}$ terms in Eq. \eqref{eq:muretro} can be dropped. The ratio between magnifications \eqref{eq:muretro} of retro-lensing images of successive orders becomes
\iea{
  \frac{\mu'_{n+1}}{\mu'_n}&\approx& \frac{c'_2(n+1)\theta'_{n+1,0}/\beta'}{c'_2(n)\theta'_{n,0}/\beta'}
  \approx e^{-2\sqrt{2}\pi\sqrt{\frac{x^2-1}{x^2+2}}}<1.\label{eq:murl2}
}
We see that these magnifications also form a geometric series, and they have the same common factor as those of strong regular lensing given by Eq. \eqref{eq:musl2}. This immediately implies that the increase of charge will have the same effect on the magnification here, i.e., increasing this common factor.

One can also compare the magnification of retro- and strong regular lensing images of the same order. Their ratio is given by the following if $\beta=\beta'$
\iea{
  \frac{\mu'_{n,\pm}}{\mu_{n,\pm}}&\approx&\frac{|c'_{2}|\theta'_{n,0}/\beta'}{c_2\theta_{n,0}/\beta} \approx \frac{|d_{\rm ls}-d_{\rm ol}|}{d_{\rm ls}+d_{\rm ol}}e^{\sqrt{2}\pi\sqrt{\frac{x^2-1}{x^2+2}}}.\label{slwlr}
}
For $d_{\rm os}$ and $d_{\rm ls}$ that are not too close such that $|d_{\rm ls}-d_{\rm ol}|/(d_{\rm ls}+d_{\rm ol})\approx 1$,
\iea{
  \frac{\mu'_{n,\pm}}{\mu_{n,\pm}}\approx e^{\sqrt{2}\pi\sqrt{\frac{x^2-1}{x^2+2}}},
}
which is always larger than 1 for all $x$. This means that the magnification of retro-lensed image is always larger than that of strong regular lensing image of the same order $n$ when taking small $\beta'=\beta$.

The velocity correction to the apparent angle \eqref{eq:thetav1retro} for the retro-lensing is given by
\iea{
	\theta'_{n,\pm}(\beta',v)&=& \pm\left\{\theta '_{n,0}(h,v)+\frac{M\sqrt{x^2-1}(x^2+2)(d_{\rm ls}-d_{\rm ol})\xi'_{n}(h,v)\left[\pm\beta' -\frac{d_{\rm ls}+d_{\rm ol}}{d_{\rm ls}-d_{\rm ol}} \theta '_{n,0}(h,v)\right]  }{x  d_{\rm ol} d_{\rm ls}(1+a_2)v}\right\},\label{eq:thetaretrov1correc}\\
&&(n=1,~2,~3,\cdots), \nonumber
}
with $a_2$, $\theta'_{n,0}(h,v)$ and $\xi'_{n}(h,v)$ given by equations \eqref{eq:defa2}-\eqref{eq:xinv} but with $(2n+1)\pi$ replaced by $2n\pi$.

\begin{figure}[htp]
  \centering
  \subfloat[]{\label{fig:thetaretroposnegv1q0n1}\includegraphics[width=0.45\textwidth]{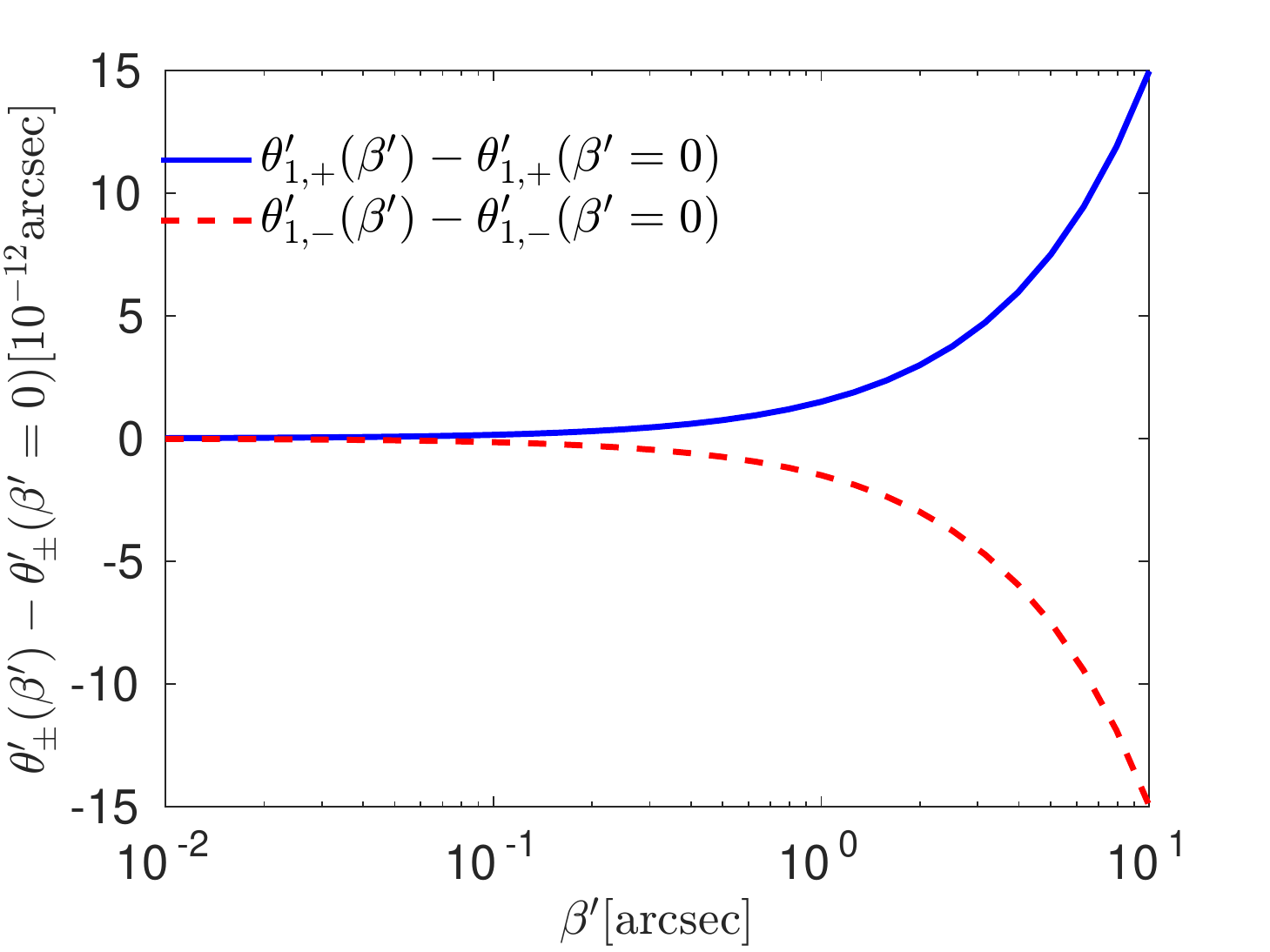}}\\
  \subfloat[]{\label{fig:thetaretroposnegv1q0beta10}\includegraphics[width=0.45\textwidth]{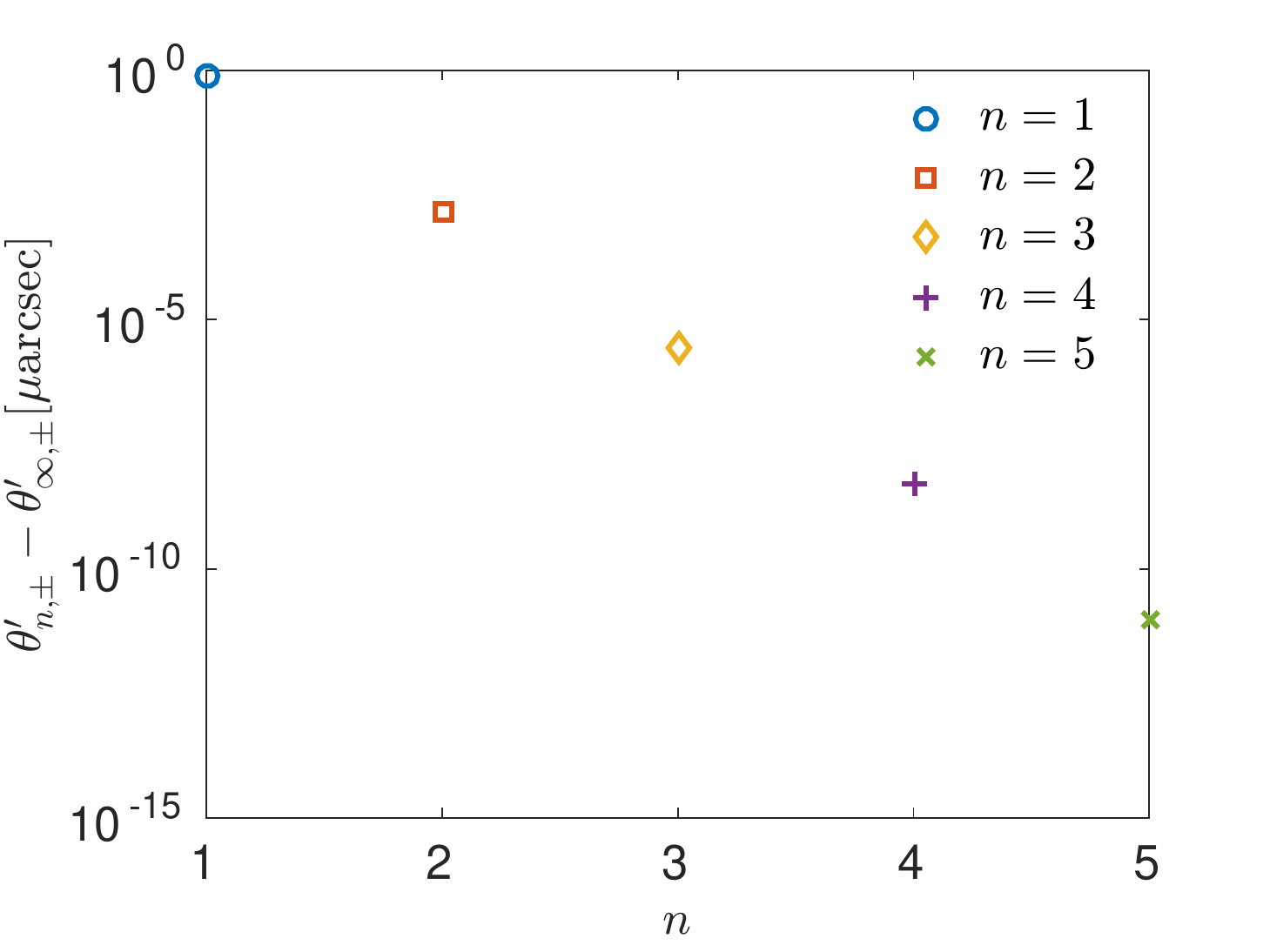}}~
  \subfloat[]{\label{fig:thetaretroposbeta10n1}\includegraphics[width=0.45\textwidth]{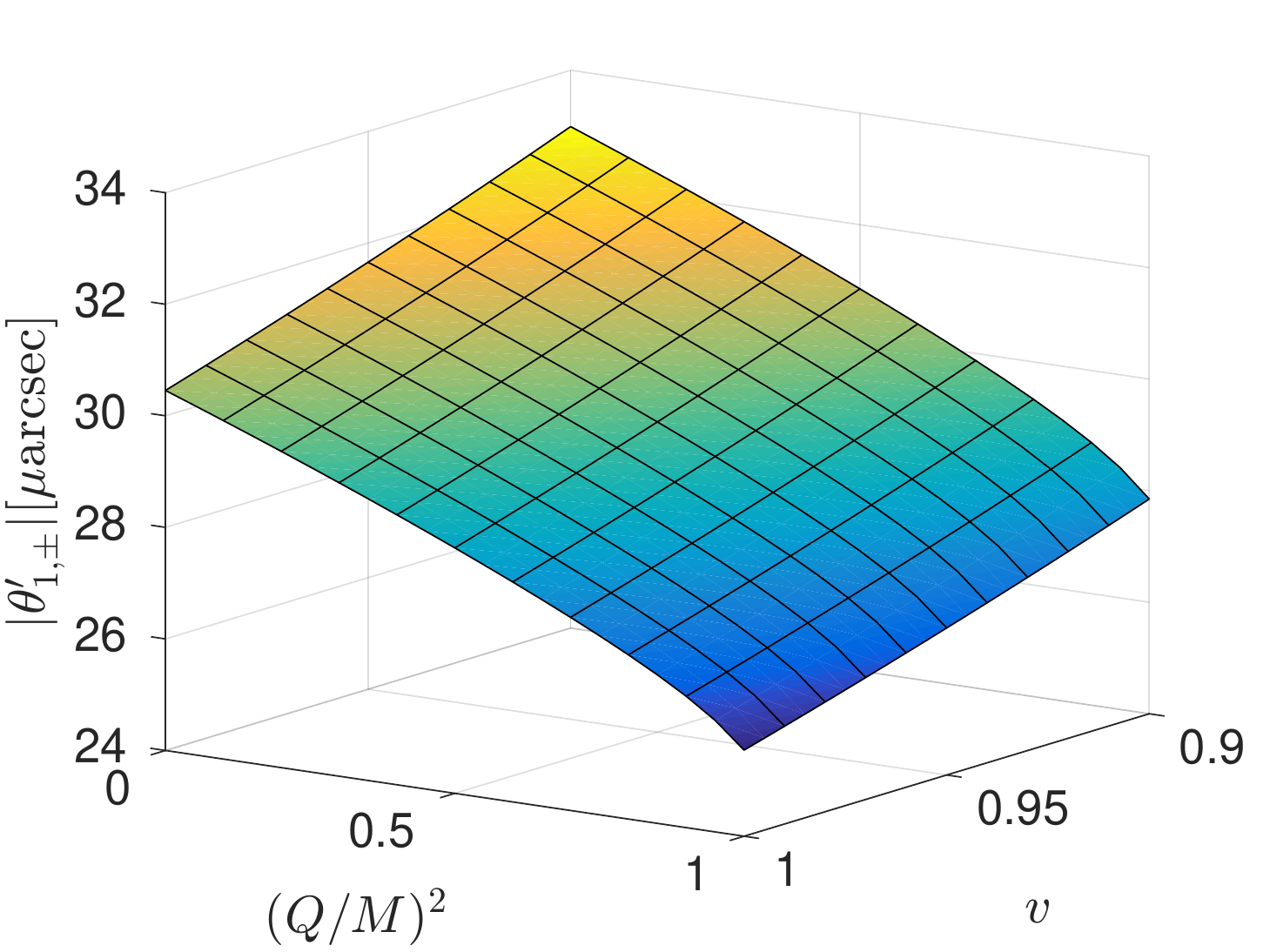}}\\
  \caption{Apparent angle of relativistic retro-lensed images.  Fig. \ref{fig:thetaretroposnegv1q0n1}: $\theta_{1,\pm}(\beta',h=0,v=1)$, where $|\theta'_{1,\pm}(\beta'=0)|=27.3678~\mathrm{\mu arcsec}$; Fig. \ref{fig:thetaretroposnegv1q0beta10}: $\theta_{n,\pm}(\beta'=10~{\rm arcsec},~h=0,~v=1)$; Fig. \ref{fig:thetaretroposbeta10n1}: $\theta_{1,\pm}(\beta=10~{\rm arcsec},~h,~v)$.  The $\theta_{n,\pm}$ in Fig. \ref{fig:thetaretroposnegv1q0beta10} and \ref{fig:thetaretroposbeta10n1} are numerically so close that they do not distinguish. Other parameters are the same as in Fig. \ref{fig:thetaweak}.}
  \label{fig:thetaretro}
\end{figure}

In Fig. \ref{fig:thetaretro}, we plot the apparent angle in the retro-lensing case as functions of $\beta', ~h,~v$ and $n$. It is seen that the apparent angle for $n=1$ is much larger than the strong lensing apparent angle with same $n$ and other parameters (see Fig. \ref{fig:thetastrongposnegv1q0n1}) but still much smaller than the weakly lensed image (see Fig. \ref{fig:thetaweak}).
Similar to the strong regularly lensed images, from Fig. \ref{fig:thetaretroposnegv1q0beta10} we also verify that indeed the asymptotic value for $\displaystyle \lim_{n\to\infty}\theta'_{n,\pm}(\beta,v)$ is numerically the same as in Eq. \eqref{eq:thetastrongv1Eninf}. The value of these apparent angles for $n$ from 1 to 5 are also listed in Table \ref{tab:thetaforn}. Finally, the effects of $Q$ and $v$ in the retro-lensing case are very similar to these in the strong regular case, both qualitatively and quantitatively.

Finally, from the analysis of the strong regular lensing in subsection \ref{sec:strongregular} and retro-lensing in subsection \ref{sec:strongretro}, we understand that these two lensing scenarios are different essentially because of the different amount of trajectory deflection. From the deflection angle point of view, the $\Delta\varphi$'s of these two kinds of lensing are staggered if arranged from small to large: any $\Delta\varphi$ at certain order from one of the lensing's should be between two $\Delta\varphi$'s at same or nearby order from the other lensing. This pattern clearly has an influence on the form of the magnification for these two kinds of lensing's. This can be seen from the geometric series
Eqs. \eqref{eq:musl2} and \eqref{eq:musl2} and the relations \eqref{slwlr}: the magnifications of the two kinds of lensing's are also staggered. Now we would like to show that this is staggering also happens to the apparent angles of the two lensing's. In Table \ref{tab:thetaforn}, we list $\theta_{n,\pm}(\beta,v)$ and $\theta'_{n,\pm}(\beta,v)$ for $\beta=\beta'=10~{\rm arcsec},~h=0$ and $v=1$. They are numerically very close to the limiting apparent angle $\theta_{\infty,\pm}$ and converge to it as $n$ increases. Therefore, to make the difference clear, we subtracted this quantity in all the apparent angles. We see from the apparent angles of the strong regular (second column) and retro-lensing (third column) with a $+$ sign, they are arranged in the order
\be
\theta'_{1,+}>\theta_{1,+}>\cdots>\theta'_{n,+}>\theta_{n,+}>\theta'_{n+1,+}>\theta_{n+1,+}>\cdots. \ee
And for apparent angles of the strong regular (fourth column) and retro-lensing (fifth column) with a $-$ sign, they are arranged in the order
\be
\theta'_{1,-}>\theta_{1,-}>\cdots>\theta'_{n,-}>\theta_{n,-}>\theta'_{n+1,-}>\theta_{n+1,-}>\cdots. \ee

\begin{table}[htp]
  \caption{The apparent angles $\theta_{n,\pm}(\beta,v)$ for strong regular lensing and $\theta'_{n,\pm}(\beta,v)$ for retro-lensing subtracted by $|\theta_{\infty,\pm}|=|\theta'_{\infty,\pm}|\approx 26.5975 ~\mu$arcsec. Other parameters are same as Fig. \ref{fig:thetaweak}. All angles differences have unit $\mu$arcsec.}
  \label{tab:thetaforn}
  \begin{tabular}{c|c|c||c|c}
	\Xhline{1.2pt}
	$n$ & $\theta_{n,+}-\theta_{\infty,+}$ & $\theta'_{n,+}-\theta'_{\infty,+}$ &$\theta_{n,-}-\theta_{\infty,-}$ &$\theta'_{n,-}-\theta'_{\infty,-}$  \\
	\Xhline{1.2pt}
	\multirow{2}{*}{$1$} && $3.86\times10^{0}$ && $2.32\times10^{0}$ \\
	&$5.66\times10^{-1}$ && $4.99\times10^{-1}$ &\\
	\hline
	\multirow{2}{*}{$2$} && $7.20\times10^{-3}$ && $4.33\times10^{-3}$ \\
	&$1.06\times10^{-3}$ && $9.32\times10^{-4}$ &\\
	\hline
	\multirow{2}{*}{$3$} && $1.35\times10^{-5}$ && $8.08\times10^{-6}$ \\
	&$1.97\times10^{-6}$ && $1.74\times10^{-6}$ &\\
	\hline
	\multirow{2}{*}{$4$} && $2.51\times10^{-8}$ && $1.51\times10^{-8}$ \\
	&$3.68\times10^{-9}$ && $3.25\times10^{-9}$ &\\
	\hline
	\multirow{2}{*}{$5$} && $4.69\times10^{-11}$ && $2.82\times10^{-11}$ \\
	&$6.88\times10^{-12}$ && $6.07\times10^{-12}$ &\\
	\Xhline{1.2pt}
  \end{tabular}
\end{table}

\section{Discussions} \label{sec:discussions}

We studied the deflection angle of lightlike and timelike particle rays in RN spacetime. It is found that this angle is expressible for any particle velocity $v$ and spacetime charge $Q$ formally as an elliptical function, as given in Eq. \eqref{eq:deltaphi}. In order for the particle to escape to spatial infinity, we found that for any $v$ and $Q$ there exist a critical impact parameter $b_c$ given by Eq. \eqref{eq:defbc}, which corresponds to a critical closest radius $r_c$ in Eq. \eqref{eq:defrc}. This $r_c$ is indeed the radius of the particle sphere for particles with velocity $v$ in RN spacetime with charge $Q$. For any fixed velocity $v$, the increase of charge will decrease both $b_c$ and $r_c$ finitely. For any fixed charge $Q$, the decrease of velocity will increase $b_c$ infinitely and $r_c$ finitely.

In order to study the effect of velocity and charge on the deflection angle, its expansions in the large and small velocity and/or charge limits are found in the weak and strong field cases. For the purpose of easier reference, these expansions are being summarized in Table \ref{tab:deltaphiexp}.
In general, we found that in all limits and also the general case, the deflection angle decreases as $Q$ increases from 0 in the Schwarzschild case to 1 in the extremal RN case and will increase infinitely as $v$ decreases from light speed to 0. For weak deflection limit, the change of the deflection angle cause by variation of velocity is at the order of $\mathcal{O} (1/b)$ while that cause by variation of charge is at the order of $\mathcal{O} (1/b^2)$ and therefore the former is much larger than the later. However, for strong field and relativistic particle limit, the effect of velocity and deflection angle are of the same $\mathcal{O} (1/\delta)$ order.

\begin{table}
\caption{The deflection angles and their expansions in various limits}
\label{tab:deltaphiexp}
\centering
\begin{tabular}{l|l|l}
  \Xhline{1.2pt}
  \multirow{4}{*}{Weak}
  &\multirow{2}{*}{Relativistic particles:  \eqref{eq:deltaphiv1finfexpv1vcorre}}
  &Small $Q$: \eqref{eq:deltaphiv1finfexpv1vcorre}\\
  \cline{3-3}
  &&Large $Q$: \eqref{eq:deltaphiv1finfexpv1vcorre}\\
  \cline{2-3}
  &\multirow{2}{*}{Non-relativistic particles: \eqref{eq:deltaphiv0finfexp1} and \eqref{eq:deltaphiv0finfexp2}}
  &Small  $Q$: \eqref{eq:deltaphiv0finfexp1} and \eqref{eq:deltaphiv0finfexp2}\\
  \cline{3-3}
  &&Large $Q$: \eqref{eq:deltaphiv0finfexp1} and \eqref{eq:deltaphiv0finfexp2}\\
  \cline{2-3}
  \hhline{=|=|=}
  \multirow{4}{*}{Strong}
  &\multirow{2}{*}{Relativistic particles: \eqref{eq:deltaphiv1fcexp}}
  &Small $Q$: \eqref{eq:deltaphiv1fcexpvcorrech0exp}\\
  \cline{3-3}
  &&Large $Q$: \eqref{eq:deltaphiv1fcexpvcorrech1exp}\\
  \cline{2-3}
  &\multirow{2}{*}{Non-relativistic particles: \eqref{eq:deltaphifcv0expomegac}}
  &Small  $Q$: \eqref{eq:deltaphifcv0expomegach0expfc4}\\
  \cline{3-3}
  &&Large $Q$: \eqref{eq:deltaphifcv0expomegach1expfch1}\\
  \Xhline{1.2pt}
\end{tabular}
\end{table}

The deflection angles are applied to the GL in RN spacetime. For the regular lensing's, including weak regular lensing and strong regular lensing, and the retro-lensing, we have solved the lens equations and obtained the apparent angles. In general, one finds that in the all of the lensing scenarios, comparing to light lensed in Schwarzschild spacetime, the velocity decrease tends to increase the apparent angles while the charge increase has an opposite effect. Both these two effects can be understood from their effects on the deflection angle found before. Moreover, for weak lensing the effect of velocity is a few orders higher than that of charge while in strong lensing  (both regular and retro) their effects are comparable. Again, this can be understood from the influence of these two factors on the deflection angle in strong field limit.

The change of apparent angle due to velocity was correlated to the neutrino mass and mass hierarchy in Ref.  \cite{Jia:2015zon}. Now if the central mass carries charge, the corresponding angular difference should be modified to
\iea{
  \theta_{\pm,\nu_i}-\theta_{\pm}|_{v=1}&=& \pm \frac{m_i^2}{E^2} (d_1+d'_1\beta),\label{eq:neutrinomlight}\\
  \theta_{\pm,\nu_1}-\theta_{\pm,\nu_2}&=& \pm \frac{m_1^2-m_2^2}{E^2} (d_1+d'_1\beta),\label{eq:neutrinomneutrino}
}
where $d_1$ and $d'_1$ are given by Eqs. \eqref{eq:defd1} and \eqref{eq:defd1p}. For the Sgr A* and a source on the edge of the galaxy stellar, $d_1$ is at the order of arcsec. The small ratio between neutrino mass square (difference) and energy square, however, highly suppresses the differences between apparent angles, making the resolution very difficult. The effect of a nonzero charge, although in general is to further reduces these two differences, is indeed numerically very small. Therefore only for more exotic and heavier particles, this angular separation might be of practial use.

One more application of our result is to constraint RN black hole charge using its shadow size caused by lensing of lightlike or timelike particles. We showed that both the strong regular lensing and retro-lensing lead to the same asymptotic apparent angle $\theta_{\infty,\pm}$ given by Eq. \eqref{eq:thetastrongv1Eninfvcorrec}, whose value defines the shadow size of the corresponding black hole. For the Sgr A* in the galactic center, using its mass $4.31\times 10^6 M_\odot$ and the distance $d_{\rm ol}=8.33$ kpc, we can directly use Eq. \eqref{eq:thetastrongv1Eninfvcorrec} to estimate its shadow size as
\be
\theta_{\infty,+}-\theta_{\infty,-}=2\theta_{\infty,+}= 53.2 p(x) ~[\mu{\rm arcsec}] + 35.5q(x)(1-v) ~[\mu{\rm arcsec}] \label{eq:szv}
\ee
where
\be p(x)=\frac{(2+x^2)^{3/2}}{3\sqrt{6}x},~q(x)=\frac{\sqrt{x^2+2}(x^2+4)}{4\sqrt{6}x}
\ee
and $x$ was in \eqref{eq:xdef}.
For lightray, we see that only the first term contributes. In this case, for Schwarzschild spacetime, $x=2$, $p(x)=1$ and the corresponding shadow size is about 53.2 $\mu{\rm arcsec}$. While for extremal RN spacetime, $x=\sqrt{2}$ and $p(x)=4/(3\sqrt{3})$, which lead to a shadow size of 40.9 $\mu{\rm arcsec}$. These are in agreement with Ref. \cite{Zakharov:2011zz} where only shadow due to lightray is studied. Now if the observed ray is not lightlike and its velocity deviate noticeably from light speed, then from Eq. \eqref{eq:szv} it is clear that this shadow size will receive a large and positive correction. For example, if $v=0.9$, then we have
\be
\theta_{\infty,+}-\theta_{\infty,-}= 53.2 p(x) ~[\mu{\rm arcsec}] + 3.55q(x) ~[\mu{\rm arcsec}]
\ee
which is 56.8 $\mu{\rm arcsec}$ for Schwarzschild spacetime and 44.0 $\mu{\rm arcsec}$ for extremal RN spacetime. The increase of these shadow sizes due to velocity makes the measurement of the shadows easier and therefore of practical use.

A few remarks regarding the possible extensions of the current work are in order.
First of all, throughout the paper, we have concentrated on the $0\leq Q\leq M$ case in order for the RN black hole to exist. However, in principle the deflection angle and apparent angles  (in some limits) we found above, are still usable for $Q>1$. For this case, it was argued in Ref. \cite{Pugliese:2011py} that the GL might be used to distinguish a black hole and a naked singularity. Although the effect of particle velocity and charge in this case might be of certain theoretical value, the existence of such spacetime are more hyper-theoretical and therefore not studied here.
Secondly, in this work we have assumed that the deflected and lensed ray are neutral particles, such as photons or neutrinos. However, in principle one can also study how the charged particle with different charge sign and value will experience the trajectory deflection. Study of this is of less usefulness in practical GL because of the short scattering length of charged particles in universe, but they might be important in processes such as accretion by RN black hole. Finally, other aspect of the particles motion in charge spacetime can also be studied, such as the effect of cosmological constant \cite{Zhao:2016ltm} or the time delay of timelike rays \cite{Zhang:2017pxr}. The time delay effect is particularly interesting because it is more realistic to observe and usual features such as negative time delays in spacetime with naked singularities \cite{Virbhadra:2007kw,DeAndrea:2014ova}. Currently, we are working along the last direction.

\begin{acknowledgments}
  The authors appreciate discussions with Mr. Chengzhe Li. This research is supported by the NNSF China 11504276 \& 11547310 and MST China 2014GB109004.
\end{acknowledgments}

\appendix

\section{Exact formula about roots and radius of particle sphere}\label{sec:exact}

The solutions of a  general quartic equation
\iea{
  a x^4+b x^3+ c x^2+d x+e=0, \label{eq:quartic}
}
are given by
\iea{
  \omega_{\substack{1\\4}}&=& -\frac{b}{4 a}+S\mp\frac{1}{2} \sqrt{-2 p-\frac{q}{S}-4 S^2},\label{eq:quarticsol14}\\
  \omega_{\substack{2\\3}}&=& -\frac{b}{4 a}-S\mp\frac{1}{2} \sqrt{-2 p+\frac{q}{S}-4 S^2},\label{eq:quarticsol23}
}
where
\iea{
  p&=& \frac{8 a c-3 b^2}{8 a^2},\label{eq:quarticp} \\
  q&=& \frac{8 a^2 d-4 a b c+b^3}{8 a^3}, \\
  S&=& \frac{1}{2} \sqrt{-\frac{2 p}{3}+\frac{2 \sqrt{\Delta _0} \cos \left (\frac{\varphi }{3}\right)}{3 a}}, \label{eq:quarticS} \\
  \varphi&=& \cos ^{-1}\left (\frac{\Delta _1}{2 \sqrt{\Delta _0^3}}\right), \\
  \Delta_0&=& 12 a e-3 b d+c^2, \label{eq:quarticdelta0}\\
  \Delta_1&=& -72 a c e+27 a d^2+27 b^2 e-9 b c d+2 c^3.\label{eq:quarticdelta1}
}
  In this paper we only consider the quartic equation whose roots are all real, therefore $S$ can be written in the real form Eq. \eqref{eq:quarticS}.

Comparing Eq. \eqref{eq:polyomega} with \eqref{eq:quartic} we can see that
\iea{
  a&=& -f^2 h^2,~b=2f^2,~c=-f^2-h^2 (1-v^2),~d=2 (1-v^2),~e=v^2, \label{eq:polyomegapara}
}
replacing these parameters in equations \eqref{eq:quarticp}-\eqref{eq:quarticdelta1} and then in equations \eqref{eq:quarticsol14} and \eqref{eq:quarticsol23}, we will get the four roots of Eq. \eqref{eq:polyomega} in the desired order $\omega_1<0<\omega_2<\omega_3<\omega_4$.

We now show that $f$ must satisfy some conditions to ensure that all roots of Eq. \eqref{eq:quartic} are real. From Eq. \eqref{eq:polyomegapara} we see that in general $a<0$ and $e>0$ for RN spacetime. Substituting them together with other quantities in Eq. \eqref{eq:polyomegapara} into Eqs. \eqref{eq:quarticp}-\eqref{eq:quarticdelta1}  and further into the square root part in Eq. \eqref{eq:quarticsol14}, a small calculation shows that they will sufficient to ensure two real roots $\omega_1<0<\omega_4$. To force the other two roots to be real, we have to demand that the left-hand side of \eqref{eq:quartic} when treated as a function of $\omega$, has a local minimum and this minimum is less than or equal zero. To find the local minimum point of $\omega$, differentiating the left-hand side of Eq. \eqref{eq:quartic} with respect to $\omega$ we get
\iea{
  -2 f^2 h^2 \omega ^3+3 f^2 \omega ^2+\omega  \left [h^2 \left (v^2-1\right)-f^2\right]-v^2+1 &=& 0. \label{eq:polyomegadiff}
}
We then should solve $\omega$ from this equation and substitute into the left-hand side of Eq. \eqref{eq:quartic} and demand the result to be less or equal zero. The critical behavior indeed happens when this minimum {\it equals} zero. Therefore, at this critical point, essentially both the Eqs. \eqref{eq:quartic} and \eqref{eq:polyomegadiff} should be satisfied simultaneously. Solving these two equations, we obtain the critical value of $f$ in terms of  other parameters $h$ and $v$
\iea{
  f_c (h,v)&=& \frac{\sqrt{1-v^2} \sqrt{h^2 \omega _c(h,v)-1}}{\sqrt{-2 h^2 \omega _c(h,v)^3+3 \omega _c(h,v)^2-\omega _c(h,v)}},
}
where $\omega_c(h,v)$ is just the $\omega_2$ in Eq. \eqref{eq:quarticsol23} but with the following parameters
\iea{
  a'&=& h^4 (v^2-1),~b'=4h^2 (1-v^2),~c'=-4+2 (2+h^2)v^2,~d'=1-4v^2,~e'=v^2. \label{eq:omegacabcde}
}
Only when $f\geq f_c(h,v)$, the four roots of Eq. \eqref{eq:quartic} will be real.

\section{Derivation of Eqs. \eqref{eq:roots14} and \eqref{eq:roots23}}\label{sec:rootsv1pert}

We begin by expanding $f_c$ to the first order of $ (1-v)$. It is seen that $f_c(h,v)$ and $\omega_c(h,v)$ can be solved from Eqs. \eqref{eq:polyomegadiff} and \eqref{eq:polyomega}. Eliminating $\omega$ from them we get the equation that $f_c(h,v)$ should satisfy
\iea{
  \Delta=\frac{1}{27}\left(4\Delta_0^3-\Delta_1^2\right)=0, \label{eq:quarticdiscri}
}
where $\Delta$ is the discriminant of Eq. \eqref{eq:polyomega}, $\Delta_0$ and $\Delta_1$ are given by Eqs. \eqref{eq:quarticdelta0} and \eqref{eq:quarticdelta1}, and the parameters in them are given by Eq. \eqref{eq:polyomegapara}.
Then to the zeroth order of $(1-v)$ we have
\iea{
  f_c (h,v=1)&=& \frac{b_c (h,v=1)v}{M}=\frac{\left (x^2+2\right)^{3/2}}{\sqrt {2} x},\label{eq:fcv1sol}
}
where $b_c (h,v=1)$ is the critical impact parameter given in Eq. \eqref{eq:bchv1}. Assuming $v=1-\delta_v$ and formally
\iea{
  f=f_{c0}+f_{c1}\delta_v,
}
we can substitute them into Eq. \eqref{eq:quarticdiscri}. After some simplification the equation at lowest nonzero order is
\iea{
  \frac{\left (x^2-1\right)^3 \left (x^2+2\right)^7 \left (2 \sqrt{2} f_{c1} \sqrt{x^2+2}+x^3+2 x\right)}{8 x^5}=0.
}
Solving this for $f_{c1}$ and substituting back into $f$, we obtain
\iea{
  f_c&=& \frac{\left (x^2+2\right)^{3/2}}{\sqrt{2} x}-\frac{ x \sqrt{x^2+2}}{2 \sqrt{2}}\delta_v, \label{eq:fcdeltav}
}

Now we are ready to calculate the expansions of roots using method of undetermined coefficients. Using Eqs. \eqref{eq:fcdeltav} and \eqref{eq:quarticsol14} and \eqref{eq:quarticsol23}, to the leading order we find
\iea{
  \omega _{\substack{1\\4},0}&=& -\frac{2 x}{ (x\pm 2) \left (x^2+2\right)},~\omega _{\substack{2\\3},0}=\frac{2}{x^2+2}.
}
For $\omega_1$, the calculation is straight forward. Letting
\iea{
v&=&1-\delta_v,~f=f_{c0}+f_{c1}\delta_v+\delta_f, \nonumber \\
\omega_1&=&\omega_{1,0}+\omega_{1,v}\delta_v+\omega_{1,f}\delta_f,\nonumber
}
in Eq. \eqref{eq:polyomega}, and comparing the coefficients of the powers of $\delta_v$ and $\delta_f$ on the two sides of the equation, we find to the lowest non-trivial order, i.e., $\delta_f^1$ and $\delta_v^1$, that
\iea{
  0&=& 4 \sqrt{x^2+2} \left (x^6+2 x^5+8 x+4\right) \omega _{1,f}-x^2 \left[x^2 \left (\sqrt{2}-20 \sqrt{x^2+2} \omega _{1,f}\right)\right.\nonumber\\
  &&-4 x \left.\left (\sqrt{2}-8 \sqrt{x^2+2} \omega _{1,f}\right) -4 \left (\sqrt{2}-8 \sqrt{x^2+2} \omega _{1,f}\right)\right], \\
  0&=&2  (x+1) \left (x^2+2\right)^2 \omega _{1,v}-x  (x+2)^2.
}
The coefficients $\omega_{1,v},~\omega_{1,f}$ can be solved from these equations, and their solutions are given in corresponding terms in  \eqref{eq:roots14}. $\omega_4$ can be handled similarly and the result is also given in Eq. \eqref{eq:roots14}. $\omega_2$ and $\omega_3$ are doubly degenerate roots, in which the $\sqrt{\delta_v}$ and $\sqrt{\delta_f}$ order terms may appear  \cite{Hinch:1991pm}. Therefore, we assume
\iea{
  \omega_2&=&\omega_{2,0}+\omega_{2,vh}\sqrt{\delta_v}+\omega_{2,fh}\sqrt{\delta_f}+\omega_{2,v}\delta_v+\omega_{2,f}\delta_f+\omega_{2,vhfh}\sqrt{\delta_v}\sqrt{\delta_f}\nonumber \\
  &&+\omega_{2,vhf}\sqrt{\delta_v}\delta_f +\omega_{2,vfh}\delta_v\sqrt{\delta_f}+\omega_{2,vf}\delta_v\delta_f.\nonumber
}
To the lowest nontrivial orders, we find
\iea{
  &&0= x^2 \left (x^2+3\right) \omega _{2,fh}^2-\frac{4 \omega _{2,fh}^2}{x^2}+\frac{2 \sqrt{2} x}{\left (x^2+2\right)^{3/2}},\\
  &&0= \omega _{2,fh} \omega _{2,vh}.
}
From these, $\omega_{2,vh}$ and $\omega_{2,fh}$ can be solved. We can work out every coefficient term by term and the final result is given in Eq. \eqref{eq:roots23}. The expansion of $\omega_3$ similarly solved and the solution is also present in Eq. \eqref{eq:roots23}.
\section{Convention of elliptic functions}\label{sec:elliptic}

\iea{
  F (\varphi|m)&=&\int_0^\varphi\frac{\dd \theta}{\sqrt{ (1-m\sin^2 (\theta))}},\\
  E (\varphi|m)&=&\int_0^\varphi\sqrt{1-m\sin^2 (\theta)}\dd \theta,\\
  K (k)&=&F\left (\left.\frac{\pi}{2}\right|k\right).
}

\bibliographystyle{apsrev4-1}
\bibliography{./references,./referencesadd,./referencesrevised}

\end{document}